\documentclass[conference,twoside]{IEEEtran} 
\IEEEoverridecommandlockouts
\pdfoutput=1
\def\xa{1.0}
\def\xb{1.0}

\usepackage{graphicx}
\usepackage[latin1]{inputenc}
\usepackage[caption=false]{subfig}
\usepackage{amsmath,amssymb,amsthm,latexsym}
\usepackage[hang,flushmargin]{footmisc}
\usepackage{algorithm}
\usepackage{algpseudocode}
\usepackage{xspace}
\usepackage[svgnames]{xcolor}
\usepackage{cite}
\usepackage{bm}
\usepackage{calc}
\usepackage{pdfpages}
\usepackage{hyperref}
 
\newtheorem{theorem}{Theorem}
\newtheorem{lemma}[theorem]{Lemma}

\newtheorem{corollary}[theorem]{Corollary}

\def\changed{\color{black}}
\def\endchanged{\color{black}}

\DeclareMathOperator{\lcm}{lcm}

\def\zin{\mathrm{in}}
\def\zout{\mathrm{out}}
\def\zcoor{\mathrm{coor}}
\def\MC{\mathcal{C}}
\def\MG{\mathcal{G}} 
\def\MM{\mathcal{M}}
\def\MB{\frac{\MM}{k}}

\def\matdesc#1(#2){
\left[\begin{array}{c}#1\end{array}\right]#2
}

\hypersetup{%
pdftitle={Repairing Multiple Failures with Coordinated and Adaptive Regenerating Codes},
pdfauthor={ Anne-Marie Kermarrec, Nicolas Le Scouarnec and Gilles Straub},
pdfkeywords={network coding, regenerating codes, threshold, self-healing, storage, exact repair, MBCR, MSCR, CROSS, coordinated, cooperative, collaborative, collective, adaptive, system},
pdfstartview={FitH},
urlcolor=darkblue,
pdfborder=0 0 0,
bookmarksnumbered
}%

\begin{document}
\sloppy

\thispagestyle{empty}

%

\title{Repairing Multiple Failures with Coordinated and Adaptive Regenerating Codes
\thanks{\noindent{}This paper was presented in part at the International Symposium on Network Coding in 2011 (NetCod'2011) at Beijing, China~\cite{NetCod2011}. It also initially appeared (September 2010) as an INRIA Research Report (http://hal.inria.fr/inria-00516647) entitled \emph{Beyond Regenerating Codes}. 
The main additions in this update (September 2013) are \emph{(i)} an expanded section on Adaptive Regenerating Codes explaining that they make no sense at the MBR point, and discussing their implementation (Section~\ref{sec:adapt}); \emph{(ii)} a section studying the impact of lazy repairs on both network repair cost but also on disk-related repair costs (Section~\ref{lr-io}); \emph{(iii)} a discussion of the related work (Section~\ref{sec:rel}).}
\thanks{
The following notice apply to the conference article published at NetCod 2011.
\copyright 2011 IEEE. Personal use of this material is permitted. Permission from IEEE must be obtained for all other uses, in any current or future media, including reprinting/republishing this material for advertising or promotional purposes, creating new collective works, for resale or redistribution to servers or lists, or reuse of any copyrighted component of this work in other works. 
}
}

\author{%
\IEEEauthorblockN{Anne-Marie Kermarrec\IEEEauthorrefmark{1}, Nicolas Le Scouarnec\IEEEauthorrefmark{2} and Gilles Straub\IEEEauthorrefmark{2}}
\IEEEauthorblockA{\IEEEauthorrefmark{1} INRIA Rennes - Bretagne-Atlantique, Rennes, France \\
                               \IEEEauthorrefmark{2} Technicolor, Rennes, France\\
                               Anne-Marie.Kermarrec@inria.fr, Nicolas.Le-Scouarnec@technicolor.com, Gilles.Straub@technicolor.com
}}

\maketitle%
\thispagestyle{empty}%
\begin{abstract}
Erasure correcting codes are widely used to ensure data persistence in distributed storage systems. This paper addresses the simultaneous repair of multiple failures in such codes. We go beyond existing work (\textit{i.e.}, regenerating codes by Dimakis \emph{et al.}) by describing  \emph{(i)} \emph{coordinated regenerating codes} (also known as \emph{cooperative regenerating codes}) which support the simultaneous repair of multiple devices, and \emph{(ii)} \emph{adaptive regenerating codes} which allow adapting the parameters at each repair. Similarly to regenerating codes by Dimakis \emph{et al.}, these codes achieve the optimal tradeoff between storage and the repair bandwidth. Based on these extended regenerating codes, we study the impact of lazy repairs applied to regenerating codes and conclude that lazy repairs cannot reduce the costs in term of network bandwidth but allow reducing the disk-related costs (disk bandwidth and disk I/O). 
\end{abstract}%
%


\section{Introduction}

Over the last decade, digital information to be stored, be it scientific data, photos, videos, \textit{etc.}, has grown exponentially. Meanwhile, the widespread access to the Internet has changed behaviors: users now expect reliable storage and seamless access to their data. The combination of these factors dramatically increases the demand for large-scale distributed storage systems for backing up or sharing data. This is traditionally achieved by aggregating numerous physical devices to provide large and resilient storage~\cite{Dabek2001,Rhea2003,Ghemawat2003,Bhagwan2004}. In such systems, which are prone to disk and network failures, redundancy is the natural solution to prevent permanent data losses.  However, as failures occur, the level of redundancy decreases, potentially jeopardizing the ability to recover the original data. This requires the storage system to self-repair to go back to its healthy state (\textit{i.e.}, keep redundancy above a minimum level).

Repairing lost redundancy from remaining one is paramount for distributed storage systems. Redundancy in storage systems has been extensively implemented using erasure correcting codes~\cite{Weatherspoon2002,Lin2004,Bhagwan2004} for they enable tolerance to failures with low storage overheads. However codes came at the price of a large communication overhead, because repairing required downloading and decoding the whole file. This repair cost has a wide impact on systems since repairs are not limited to restoring data after permanent failures, but are also triggered when doing degraded reads (\emph{i.e.}, accessing data stored on temporarily unavailable or overloaded devices). Dimakis \emph{et al.} recently showed~\cite{Dimakis2007,Dimakis2010} that the repair cost can be significantly reduced by avoiding decoding using \emph{regenerating codes}.  Yet, they assume a static setting and do not support simultaneous coordinated repairs.

In this paper, we go beyond these works by considering simultaneous repairs in regenerating-like codes. We propose \emph{coordinated regenerating codes} allowing devices to leverage simultaneous repairs (or simultaneous degraded reads): each of the $t$ devices being repaired contacts $d$ live (i.e., non-failed) devices and then coordinates with the $t-1$ others. We also consider a relaxed scheme where $d$ and $t$ can change at each repair to define \emph{adaptive regenerating codes}. Our contributions regarding these codes are:%
\begin{itemize}%
\item We define \emph{coordinated regenerating codes} (also known as \emph{cooperative regenerating codes}) and derive closed form expressions of the optimal quantities of information to transfer when $t\negthinspace>\negthinspace 1$ devices must be repaired simultaneously from $d$ live devices (Section~\ref{sec:coord}).
\item We design \emph{adaptive regenerating codes} achieving optimal repairs in a dynamic environment where $t$ and $d$ change over time. (Section~\ref{sec:adapt}).
\item Based on these constructions, we prove that, when relying on regenerating-like codes (MSR or MBR)~\cite{Dimakis2010}, deliberately delaying repairs does not bring further savings with respect to repair bandwidth, contrary to what is observed for traditional erasure correcting codes~\cite{Bhagwan2004,Datta2006,Dalle2009} but that it could help when looking at disk I/O (Section~\ref{sec:lazy}).
\end{itemize}

\begin{figure*}[!t]
	\centering
	\subfloat[Erasure correcting codes]{
		\includegraphics[height=0.120\linewidth]{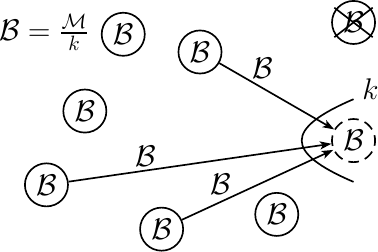}
		\label{fig:network_ecc}
        	} \hfil
	\subfloat[Erasure codes (delayed repair)]{
		\includegraphics[,height=0.120\linewidth]{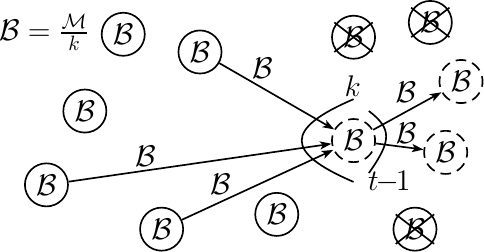}
		\label{fig:network_ecc_t}
        	} \hfil
	\subfloat[Regenerating codes]{
		\includegraphics[,height=0.120\linewidth]{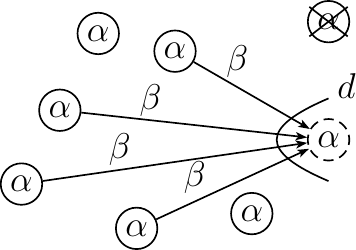}
		\label{fig:network_rc}
        	} \hfil
	\subfloat[Coordinated reg. codes]{
		\includegraphics[height=0.120\linewidth]{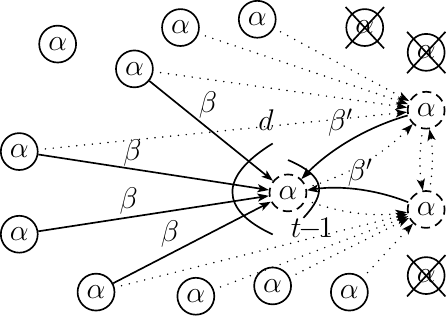}
		\label{fig:network_rc_t2}
	}
	\label{fig:network}
	\caption{Repairing failures with codes. In an $n$ device network, failed devices are replaced by new ones. The new devices fetch a given amount of data from live devices to repair the redundancy. In our examples, $k=3$, $d=4$, $t=2$ $\mathcal{B}=1$, $\alpha=1$, and $\beta=1/2$.}
		\vspace{-0.2cm}
\end{figure*}

Our work fills the gap between approaches not supporting simultaneous coordinated repair~\cite{Dimakis2010} and approaches repairing by decoding the whole file~\cite{Weatherspoon2002,Bhagwan2004,Lin2004,Datta2006,Dalle2009}. Two recent pieces of work focus on similar problems: MCR codes~\cite{Hu2010} define MSR-like codes that support multiple repairs and MFR~\cite{Wang2010} codes turn MSR codes into adaptive codes. Yet, MCR codes only consider the MSR point and assume that all transfers are equal without proving it (\textit{i.e.}, $\beta=\beta'$);  MFR~\cite{Wang2010} codes are not optimal when repairing more than one failure. More recently, concurrent studies have led to the definition of \emph{cooperative regenerating codes}~\cite{Shum2011,Shum2011b} which are similar to \emph{coordinated regenerating codes}: they also describe exact codes constructions that achieve the bounds given in this paper.

\section{Background}
\label{sec:background}

We consider an $n$ device system storing a file of $\MM$ bits split into $k$ blocks of size $\mathcal{B}=\MB$. To cope with device failures, blocks are stored with some redundancy so that a small number of failures cannot cause permanent data losses. We use a code-based redundancy scheme as it has been acknowledged as more efficient than replication with respect to both storage and repair costs~\cite{Weatherspoon2002}.  We focus on self-healing systems as they do not gradually lose their ability to recover the initial file. In the rest of this section, we describe the main code-based approaches for redundancy. For the sake of clarity we will use repairs to designate both repairs following permanent failures and degraded reads following temporary unavailability. Table~\ref{tab:example} gives some values of the storage $\alpha$ and repair $\gamma$ costs for these approaches, and also includes the codes we propose. 

\begin{table}[b]%
\caption{Some examples of repairs of codes for a file of $32$ MB\vspace{-1eX}}%
\label{tab:example}%
\centering
\begin{tabular}{|l|c|c|c|c|c|}%
\hline%
                                        		 & $k$ & $d$ & $t$ & $\alpha$ & $\gamma$ \\\hline%
Erasure codes 	               		 & 32   & \emph{NA}   & \emph{NA}    &   1 MB     &    32 MB     \\\hline
Erasure codes 	 (delayed repair)    	 & 32   & \emph{NA}   & 4    &   1 MB     &      8.8 MB  \\\hline
Dimakis \emph{et al.}'s MSR         	 & 32   & 36   & \emph{NA}    &   1 MB     &      7.2 MB \\\hline 
Dimakis \emph{et al.}'s MBR               &  32   & 36   & \emph{NA}    &   1.8 MB     & 1.8 MB \\\hline 
Our MSCR (cf. Sec. \ref{ssec:mscr})              & 32   & 36   & 4    &   1 MB     &  4.9 MB \\\hline
Our MBCR (cf. Sec. \ref{ssec:mbcr})  	 & 32   & 36   & 4    &   1.7 MB     &  1.7 MB \\\hline
\end{tabular}%
\end{table}%

\subsection{Erasure correcting codes (immediate/eager repairs)}
Erasure correcting codes have been widely used to provide redundancy in distributed storage systems~\cite{Weatherspoon2002,Lin2004}. Devices store $n$ encoded blocks of size $\mathcal{B}$, which are generated from the $k$ original blocks. The whole file can be recovered, in spite of failures, by decoding from any $k$ encoded blocks. Yet, repairing a single lost encoded block is very expensive since the device must download $k$ encoded blocks and decode the file to regenerate any single lost block (Fig.~\ref{fig:network_ecc}).

\subsection{Erasure correcting codes (delayed/lazy repairs)} 
\begin{figure}[!bh]%
	\centering%
	\subfloat[Immediate repairs]{%
		\includegraphics[width=\xb\linewidth*\real{0.38}]{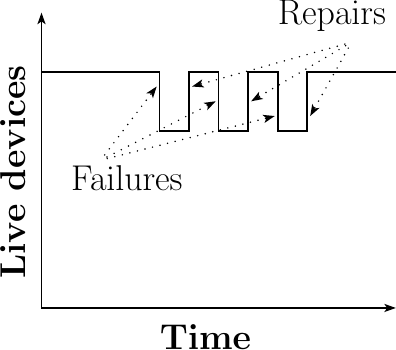}%
	   	\label{fig:nothres}%
        	} \hfil%
	\subfloat[Delayed repairs]{%
		\includegraphics[width=\xb\linewidth*\real{0.38}]{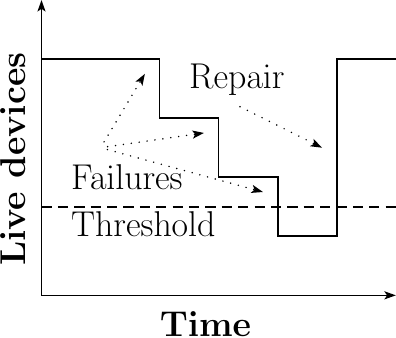}%
	   	\label{fig:thres}%
        	}%
	\caption{Delaying repairs allows performing multiple repairs at once.}%
	\label{fig:plot_thres}%
\end{figure}

A first approach to limiting the repair cost of erasure correcting codes is to delay repairs so as to factor downloading costs~\cite{Datta2006,Bhagwan2004,Dalle2009}. When a device has downloaded $k$ blocks, it can produce as many new encoded blocks as wanted without any additional cost. Hence, instead of immediately repairing every single failure (Figure~\ref{fig:nothres}), one deliberately waits until $t$ failures are detected  (Figure~\ref{fig:thres}), then one of the new devices downloads $k$ blocks, regenerates $t$ blocks and dispatches them to the $t-1$ other devices (Fig.~\ref{fig:network_ecc_t}). 

\subsection{Network coding and regenerating codes} 
A second approach to increasing the efficiency of repairs relies on network coding~\cite{Ahlswede2000}. Network coding was initially applied to multicast, for which it has been proven that linear codes achieve the maxflow in a communication graph~\cite{Li2003,Koetter2003}. Network coding has latter been applied to distributed storage and data persistence~\cite{Dimakis2005,Dimakis2006,Kamra2006,Lin2007}. A key contribution in this area is regenerating codes~\cite{Dimakis2007,Dimakis2010} introduced by Dimakis \emph{et al.}.

\begin{figure}[h]
		\centering
		\includegraphics[width=\xb\linewidth*\real{0.65}]{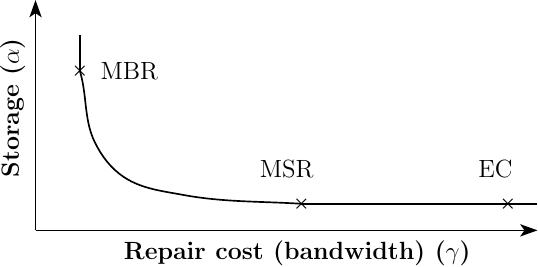}
	\caption{Regenerating codes (MSR or MBR) offer improved performances when compared to erasure correcting codes (EC)}
	\label{fig:tradeoff}
\end{figure}

Regenerating codes achieve an optimal trade-off between the storage $\alpha$ and the repair cost (repair bandwidth) $\gamma=d\beta$ with $\beta$ bits being downloaded from $d \ge k$ devices as shown on Figure~\ref{fig:network_rc}. On the tradeoff curve~(Figure~\ref{fig:tradeoff}), two specific codes are of interest: MSR (Minimum Storage Regenerating codes) which offer optimal repair costs $\gamma=\frac{\MM}{k}\frac{d}{d-k+1}$ for minimum storage costs $\alpha=\MB$ and MBR (Minimum Bandwidth Regenerating codes) which offer optimal storage costs $\alpha=\frac{\MM}{k}\frac{2}{2d-k+1}$ for minimum repair costs $\gamma=\frac{\MM}{k}\frac{2d}{2d-k+1}$. Regenerating codes can be implemented using linear codes~\cite{Li2003,Koetter2003,Ho2006,Dimakis2010b,Wu2009,Duminuco2009,Rashmi2009,Shah2010,Suh2010}.  Related work on the implementation of regenerating codes is discussed in more details in Section~\ref{sec:rel}.

\section{Coordinated regenerating codes}%
\label{sec:coord}

Regenerating codes by Dimakis \emph{et al.} perform all repairs independently. Hence, the repair cost increases linearly with $t$. In this work, we investigate repairing simultaneous failures through coordination in an attempt to reduce the cost, along the lines of delayed erasure correcting codes.  We consider that $t$ devices fail and that $t$ repairs are performed simultaneously.  
%
%

\subsection{Repair algorithm}
\label{ssec:repair}
\begin{figure}[!h]
	\centering
	\includegraphics[width=\xb\linewidth*\real{0.9}]{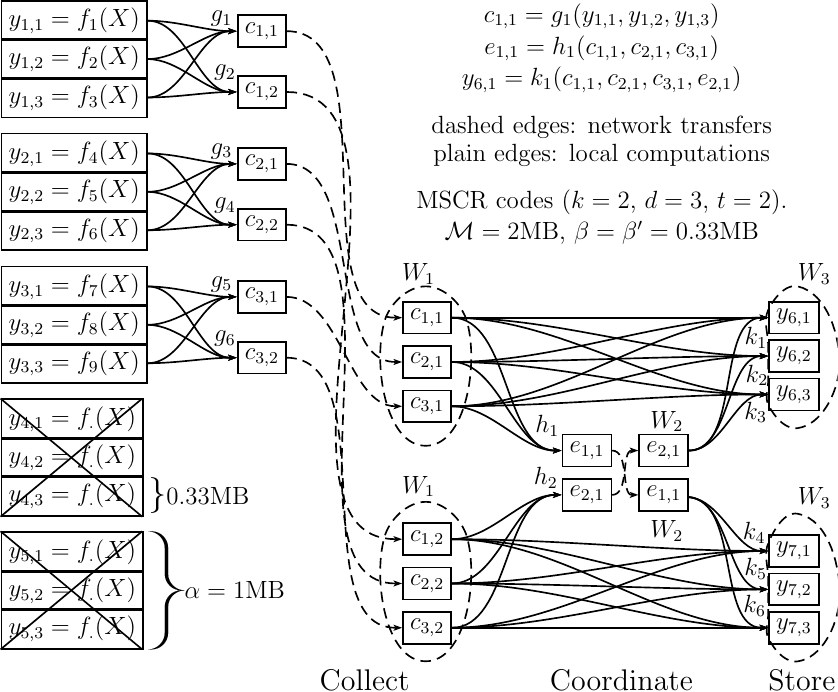}
	\caption{Coordinated regenerating codes based on linear codes. The system stores a file $X$ and is compound of 5 devices. Device $i$ stores 3 sub-blocks $\{y_{i,1},y_{i,3},y_{i,3}\}$. Devices 4 and 5 fail and are replaced by devices 6 and 7. In the figure, which depicts the block computed, $y_{i,j}, c_{i,j}, e_{i,j}$ are blocks, and $g_i, h_i, k_i$ are linear functions that compute new blocks from other blocks.\vspace{-2eX}}
	\label{fig:implem}
\end{figure}

Contrary to erasure correcting codes delayed repair (Fig.~\ref{fig:network_ecc_t}), our algorithm (Fig.~\ref{fig:network_rc_t2}) is fully distributed: repairing does not require a single device to gather all the information since no decoding is performed. A device being repaired performs the three following tasks as depicted on Figure~\ref{fig:implem}:%
\begin{IEEEdescription}[\IEEEsetlabelwidth{$\negthickspace\negthickspace\negthickspace\negthickspace$}]%
\item \textbf{1. Collect.} Download a set of sub-blocks (size $\beta$) from each of the $d$ live devices. The union of the sets is stored as $W_1$.
\item \textbf{2. Coordinate.} Upload a set of sub-blocks (size $\beta'$) to each of the $t-1$ other devices being repaired. These sets are generated from $W_1$. At this stage, sub-blocks received from the $t-1$ other devices being repaired are stored as $W_2$. 
\item \textbf{3. Store.} Store a set $W_3$ of sub-blocks (size $\alpha$) generated from $W_1 \cup W_2$. $W_1$ and $W_2$ can be erased afterwards.
\end{IEEEdescription}%
Interestingly, coordinated regenerating codes evenly balance the load on all devices, thus avoiding the bottleneck existing in erasure correcting codes delayed repairs (\textit{i.e.}, the device gathering and decoding all the information (Fig.~\ref{fig:network_ecc_t})).
 
In the rest of this section, we take an information theoretic point of view and focus on the amounts $(\alpha,\beta,\beta')$ of information exchanged.  We define the achievable tradeoffs between the storage cost $\alpha$ and the repair cost $\gamma$.

Overall, our main proof (of Theorem~\ref{thm:crc}) follows the same methodology as the seminal article by Dimakis \emph{et al.}~\cite{Dimakis2010}: the system is represented as an information flow graph, we determine inequalities on the amount of information that can flow through the graph and, applying network coding theory, we show that the recovery of a file is possible if and only if some constraints are satisfied. Costs shown in plots are normalized by $\MB$. The following table summarizes the notations used.%
\small{\begin{center}\begin{tabular}{|c|l|c|l|}%
\hline%
$k$ & Devices to recover & $\alpha$     &  Bits stored \\\hline%
$t$ & Devices being repaired & $\beta$     &  Bits transferred (collect)\\\hline%
$d$ & Live devices ($d \ge k$) & $\beta'$     & Bits transferred (coordinate)\\\hline%
$\gamma$ & \multicolumn{3}{|l|}{Total bits transferred per node repaired (\textit{i.e.}, repair cost)}\\\hline%
\end{tabular}\\[0.0eX]
\end{center}}
\normalsize%

\subsection{Information flow graphs}%
Information flow graphs describe the amounts of information transferred, processed and stored. Contrary to the graph defined in~\cite{Dimakis2010}, ours captures the coordination by adding edges between nodes being repaired. The information flow graph $\MG$ is a directed acyclic graph consisting of a source $\mathit{S}$, intermediary nodes, and data collectors $\mathit{DC}_j$ which contact $k$ devices to recover the file. A device $x^{i,j}$ is represented by $3$ nodes of the graph ($x_{\zin^{i,j}}$, $x_{\zcoor^{i,j}}$ and $x_{\zout^{i,j}}$) corresponding to its repair states ($i$ corresponds to a time step while $j$ corresponds to a device introduced at time step $i$). The capacities of the edges $(\alpha,\beta,\beta')$ correspond to the amounts of information that can be stored or transferred.

\begin{figure}[!b]
	\centering
	\includegraphics[width=\xa\linewidth*\real{0.72}]{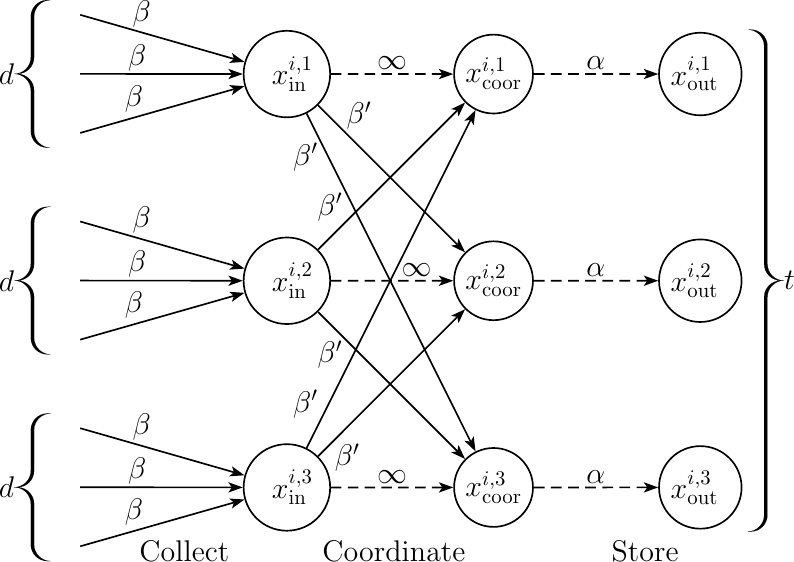}
	\caption{Information flow graph of a repair of $t=3$ devices. The internal nodes represent intermediary steps in the repair. Plain edges correspond to network communication and dashed edges correspond to local communication. }
	\label{fig:repair}
\end{figure}

Figure~\ref{fig:repair} depicts the graph of $t$ devices being repaired (assuming $t$ divides $k$.). First, devices being repaired perform a collecting step represented by $d$ edges $x_{\zout^{k,j}} \rightarrow x_{\zin^{i,j'}}$ ($k < i$) of capacity $\beta$. Second, devices undergo a coordinating step represented by $t-1$  edges $x_{\zin^{i,j}} \rightarrow x_{\zcoor^{i,j'}}$ of capacity $\beta'$ for $j \neq j'$. Devices keep everything they obtained during the first step justifying the infinite capacities of edges $x_{\zin^{i,j}} \rightarrow x_{\zcoor^{i,j}}$. Third, they store $\alpha$ as shown on edges $x_{\zcoor^{i,j}} \rightarrow x_{\zout^{i,j}}$. Figure~\ref{fig:ifg} depicts the information flow graph of successive repairs.

The graph $\MG$ evolves as repairs are performed. When a repair is performed, a set of nodes is added to the graph and the nodes corresponding to failed devices become inactive (\textit{i.e.}, data collectors and subsequently added nodes cannot connect to  these nodes). The rest of the analysis relies on the concept of maxflow, which is the maximum amount of information that can flow from the source $\mathit{S}$ to some destination $\mathit{DC}$, through the study of the minimum cut. Network coding~\cite{Ahlswede2000,Li2003,Koetter2003} allows achieving the maximum flow for multiple destinations.

\subsection{Achievable codes}

We define two important properties on codes:
\begin{IEEEdescription}[\IEEEsetlabelwidth{Correctnesi}]
\item[Correctness] A code $(n,k,d,t,\alpha,\gamma)$ is correct iff, for any succession of repairs, a data collector can recover the file by connecting to any $k$ devices.
\item[Optimality] A code $(n,k,d,t,\alpha,\gamma)$ is optimal iff it  is correct and  any code $(n,k,d,t,\bar{\alpha},\bar{\gamma})$ with $(\bar{\alpha},\bar{\gamma})<(\alpha,\gamma)$ is  not correct\footnote{In this paper, we always consider that $(\bar{a},\bar{b}) < (a,b)$ means that either $\bar{a} \le a$ and $\bar{b} < b$, or $\bar{a} < a$ and $\bar{b} \le b$  }.
\end{IEEEdescription}

The following theorem is an important result of our work.

\begin{theorem}
\label{thm:crc}
A coordinated regenerating code $(n,k,d,t,\alpha,\gamma)$ is correct\footnote{We assume that $t$ divides $k$, no result is known if $t$ does not divide $k$.} if and only if there exists $\beta$ and $\beta'$ such that the constraints of~\eqref{eq:obj} and~\eqref{eq:cons} are satisfied.  A code minimizing the repair cost $\gamma$~\eqref{eq:obj}, along constraints of~\eqref{eq:cons} is optimal. 
\begin{equation}%
\gamma=d\beta+(t-1)\beta'%
\label{eq:obj}%
\end{equation}%
\begin{align}%
 \forall \mathbf{u}, \textrm{ such that } \sum_{i=0}^{g-1}{u_i}=k  \textrm{ and } 1 \le u_i \le t, \notag \\%
\sum_{i=0}^{g-1} u_i \min\{\alpha,(d-\sum_{j=0}^{i-1}{u_j})\beta+(t-u_i)\beta'\} \ge \MM%
 \label{eq:cons}%
\end{align}%
\end{theorem}%
These constraints mean that for any scenario $\mathbf{u} = (u_i)_{0\le i<g}$ ($u_i$ is the number of devices contacted in each repair group of size $t$ during the recovery and $g$ is the number of such groups), the sum of the amounts of information that can be downloaded from each of the $k$ devices contacted by a data collector must be greater than the file size.  We now give the proof of this theorem. We study all possible graphs given some $d$, $k$ and $t$. Finally, it is shown that~\eqref{eq:cons} must be satisfied to allow decoding at any time thus preventing data losses. 


\begin{figure*}[!t]%
        \centering%
        \subfloat[$\MG^{\star}_{1}$ for $\mathbf{u}=(t,t,t)=(3,3,3)$]{%
                \includegraphics[width=0.48\linewidth]{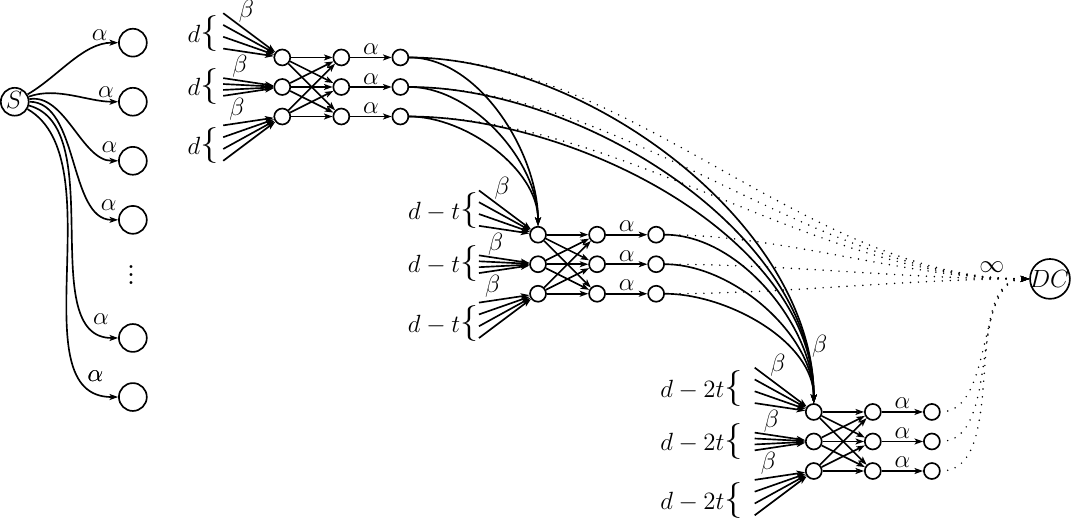}
                \label{fig:g1}%
                } \hfil %
        \subfloat[$\MG^{\star}_{2}$ for $\mathbf{u}=(1,1,1,\dots)$]{%
                \includegraphics[width=0.48\linewidth]{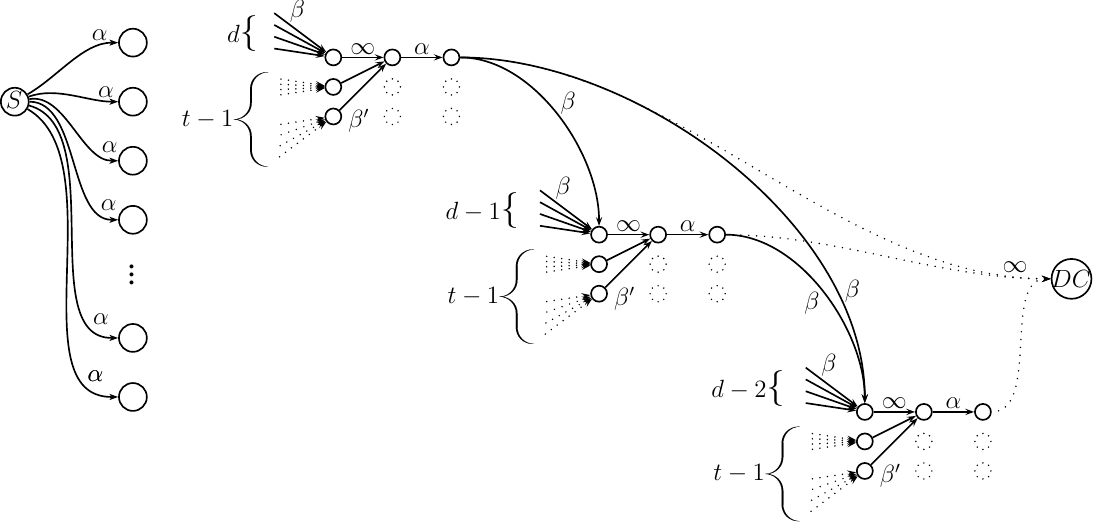}
                \label{fig:g2}%
                } \hfil%
        \subfloat[$\MG^{\star}_{3}$ for $\mathbf{u}=(2,1,3,\dots)$]{%
                \includegraphics[width=0.48\linewidth]{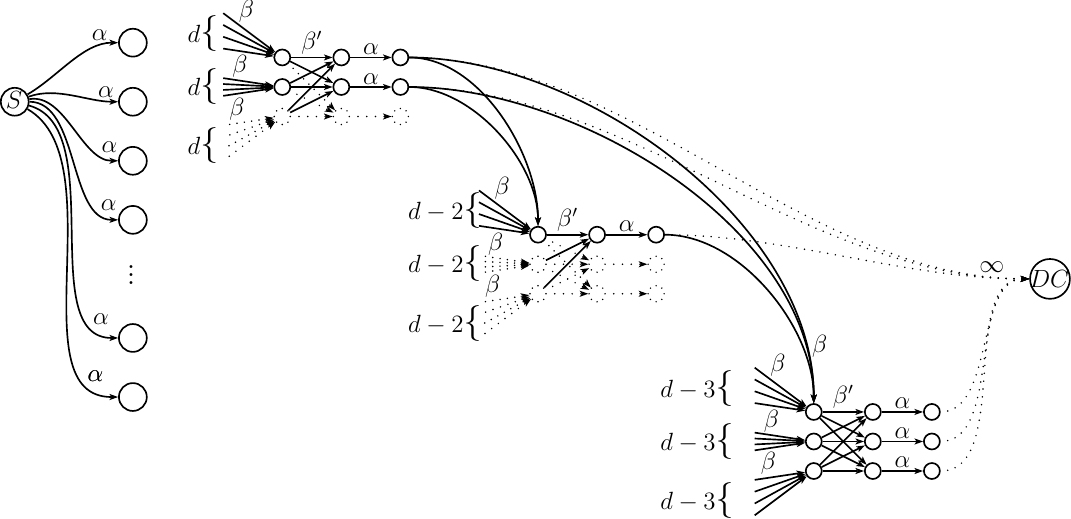}
                \label{fig:g3}%
                }%
        \caption{Information flow graphs for which bounds in~\eqref{eq:cons} are matched with equality for some $\mathbf{u}$.}%
        \label{fig:ifg}%
\end{figure*}%


\begin{lemma}
For any information flow graph $\MG$, no data collector $\mathrm{DC}$ can recover the initial file if the minimum cut in $\MG$ between $\mathrm{S}$ and $\mathrm{DC}$ is smaller than the initial file size $\MM$.
\label{lem:info}%
\end{lemma}%
\begin{IEEEproof}
Similarly to the proof in~\cite{Dimakis2010}, since each edge in the information flow graph can be used at most once, and since source to data collector capacity is less than the file size $\MM$, the recovery of the file is impossible.
\end{IEEEproof}

\begin{lemma}%
For any finite information flow graph $\mathcal{G}$, if the minimum of the min-cuts separating the source and each data collector is larger than or equal to the file size $\MM$, then there exists a linear network code such that all data collectors can recover the file. We also assume that the finite field size is not an issue.
\label{lem:info2}
\end{lemma}%
\begin{IEEEproof}%
Similarly to the proof in~\cite{Dimakis2010}, since the reconstruction problem reduces to multicasting on all possible data collectors, the result follows from network coding theory. 
\end{IEEEproof}
%
%
%
\begin{lemma}%
For any information flow graph $\MG$ consisting of initial devices that obtain $\alpha$ bits directly from the source $\mathit{S}$ and of additional devices that join the graph in groups of $t$ devices obtaining $\beta$ from  $d$ existing devices and $\beta'$ from each of the other $t-1$ joining devices, any data collector $\mathit{DC}$ that connects to a subset of $k$ out-nodes of $\MG$ satisfies:
\begin{gather}%
\mathrm{mincut}(\mathit{S},\mathit{DC}) \ge  \\\notag
\min_{\mathbf{u} \in P}\left(\sum_{i=0}^{g-1} u_i \min\{\alpha,(d-\sum_{j=0}^{i-1}{u_j})\beta+(t-u_i)\beta'\}\right) 
\label{eq:lemcons}
\end{gather}
with $P = \{\mathbf{u} : 1 \le u_i \le t \land \sum_{i=0}^{g-1}{u_i}=k \}$.
\label{lem:glob}
\end{lemma}%
\begin{IEEEproof}%
Let us consider some graph $\MG$ (see an example in Figure~\ref{fig:ifg}) formed by adding devices according to the repair process described above. Consider a recovery scenario $\mathbf{u} \in P$ in which, a data collector $\mathit{DC}$ connects to a subset of $k$ nodes $\{x_{\zout}^{i,j} : (i,j) \in I\}$, where $I$ is the set of contacted devices.


As all incoming edges of $\mathit{DC}$ have infinite capacity, we only examine min-cuts $(U,\bar{U})$ with $\mathit{S} \in U$ and $\{x_{\zout}^{i,j} : (i,j) \in I\} \subset \bar{U}$. Moreover some additional cases cannot happen since there is an order between $x_{\zin^{i,j}}$, $x_{\zcoor^{i,j}}$ and $x_{\zout^{i,j}}$ (\emph{e.g.}, $x_{\zin^{i,j}} \in \bar{U}$ and $x_{\zcoor^{i,j}} \in U$ need not be considered). Therefore, we only need to examine three cases detailed in the rest of this proof.

Let $\MC$ denote the edges in the cut (\textit{i.e.}, the set of edges going from $U$ to $\bar{U}$). Every directed acyclic graph has a topological sorting, which is an ordering of its vertices such that the existence of an edge $x \rightarrow y$ implies  $x < y$. In the rest of the analysis, we group nodes that were repaired simultaneously. Since we contact $k$ nodes, we have at least $k/t$ groups and at most $k$ groups (\emph{i.e.,} $k/t \le  g  \le k$). Since nodes are sorted, nodes considered at the $i$-th step cannot depend on nodes considered at $j$-th steps with $j>i$.

Consider the $i$-th group. Let  $J_i$ be the set of indexes such that $\{x_{\zout^{i,j}} : j \in J_i\}$ are the topologically $i$-th output nodes in $\bar{U}$ corresponding to the $i$-th (same) repair. The set contains $\#\{x_{\zout^{i,j}} : j \in J_i\}=u_i$ nodes. Consider a subset $M \subset  J_i$ of size $m$ such that $\{x_{\zin^{i,j}} : j \in M\} \subset U$ and $\{x_{\zin^{i,j}} : j \in J_i-M\} \subset \bar{U}$. $m$ can take any value between $0$ and $u_i$. 

First, consider the $m$ nodes $\{x_{\zin^{i,j}} : j \in M\}$. For each node, $x_{\zin^{i,j}} \in U$. We consider the two cases.
\begin{itemize}
\item If $x_{\zcoor^{i,j}} \in  U$,
          then $x_{\zcoor^{i,j}} \rightarrow x_{\zout^{i,j}} \in \MC$.
          The contribution to the cut is $\alpha$.
\item If $x_{\zcoor^{i,j}} \in \bar{U}$, 
         then $x_{\zin^{i,j}} \rightarrow x_{\zcoor^{i,j}} \in \MC$.
         The contribution to the cut is $\infty$.
\end{itemize}

Second, consider the $u_i-m$ other nodes $\{x_{\zin^{i,j}} : j \in J_i-M\}$ (third and last case: $x_{\zin^{i,j}}$, $x_{\zcoor^{i,j}}$ and $x_{\zout^{i,j}}$ all belong to $\bar{U}$). For each node, the contribution comes from multiple sources.
\begin{itemize}
\item The cut contains at least $d-\sum_{j=0}^{i-1}{u_j}$ edges carrying $\beta$: since $x_{\zout^{i,j}}$ are the topologically $i$-th output nodes in $\bar{U}$, at most $\sum_{j=0}^{i-1}{u_j}$ edges come from output nodes in $\bar{U}$, other edges come from $U$.
\item The cut contains $t-u_i+m$ edges carrying $\beta'$ thanks to the coordination step. The node $x_{\zcoor^{i,j}}$ has $t$  incoming edges $x_{\zin^{i,k}}  \rightarrow x_{\zcoor^{i,j}}$. However, since $\#(\{x_{\zin^{i,k}}\} \cap \bar{U}) = u_i-m$, the cut contains only $t-(u_i-m)$ such edges.
\end{itemize}

Therefore, the total contribution of these nodes is %
\begin{equation*}
c_i(m) \ge m\min(\alpha,\infty) + (u_i-m)((d-\sum_{j=0}^{i-1}{u_j})\beta+(t-u_i+m)\beta')
\end{equation*}
Since the function $c_i$ is concave for $m$ taking values in the interval $[0:u_i]$, the contribution can be bounded thanks to Jensen's inequality.%
\begin{equation*}
c_i(m) \ge u_i\min\{\alpha, (d-\sum_{j=0}^{i-1}{u_j})\beta+(t-u_i)\beta'\}
\end{equation*}

Summing these contributions for all $i$, and considering the worst case  for $\mathbf{u} \in P$ (\emph{i.e.}, the scenario $u$ that minimizes the sum) leads to~\eqref{eq:lemcons}.
\end{IEEEproof}

\begin{IEEEproof}[Proof of Theorem \ref{thm:crc}]%
From Lemmas~\ref{lem:info2} and~\ref{lem:glob}, a code is correct if it satisfies~\eqref{eq:obj} and~\eqref{eq:cons}. 
From Lemma~\ref{lem:info}, a code is correct only if $\mathrm{mincut}(\mathit{S},\mathit{DC}) \ge \MM$. Moreover, for any set of parameter $(n,k,d,t,\alpha,\beta,\beta')$ and any scenario $\mathbf{u}$, we can find a graph $\mathcal{G}_{\mathbf{u}}$ such that 
\begin{align*}
\mathrm{mincut}(\mathit{S},\mathit{DC})  =  \sum_{i=0}^{g-1} u_i \min\{\alpha,(d\negthinspace-\negthinspace\sum{u_j})\beta+(t\negthinspace-\negthinspace{}u_i)\beta'\}
\end{align*}
The graph $\mathcal{G}_{\mathbf{u}}$  is built using the following process (for $\mathbf{u}=[2,1,3]$  the graph of Figure~\ref{fig:g3} is built):
\begin{itemize}
\item The data collector gets all bits from a set $U$ of $k$ devices.
\item The contacted devices repaired simultaneously are grouped in subsets $U_i$ of size $u_i$ such that $U=\bigcup_{i=0}^{g-1}{U_i}$. Since we contact $k$ nodes, we have at least $k/t$ groups and at most $k$ groups (\emph{i.e.,} $k/t \le  g  \le k$).
\item Each device $x \in U_i$ gets $\beta$ bits from all devices in $\bigcup_{j=0}^{i-1}{U_j}$,  $\beta'$ from $u_i-1$ devices taking part to the reconstruction, $\beta$ from $d-\sum_{j=0}^{i-1}{u_j} $ devices not in $U$,  $\beta'$ from $t-u_i$ devices not taking part to the reconstruction.
\end{itemize}

Hence, a code is correct if and only if~\eqref{eq:obj} and~\eqref{eq:cons} are satisfied.
A code minimizing $(\alpha,\gamma)$ under constraints of~\eqref{eq:obj} and~\eqref{eq:cons} is optimal as any code with $(\bar{\alpha},\bar{\gamma}) < (\alpha,\gamma)$ would not satisfy at least one constraint and hence would not be correct.
%
%
%
\end{IEEEproof}

\subsection{Optimal tradeoffs}

Determining the optimal tradeoffs boils down to minimizing storage cost $\alpha$ and repair cost $\gamma$, under constraints of~\eqref{eq:obj} and~\eqref{eq:cons}. $\alpha$, $\beta$ and $\beta'$ are parameters to be optimized. Again, we assume that $t$ divides $k$.

\subsubsection{MBCR codes}
\label{ssec:mbcr}
\begin{figure}[!h]
	\centering
	\includegraphics[width=\xb\linewidth*\real{0.9}]{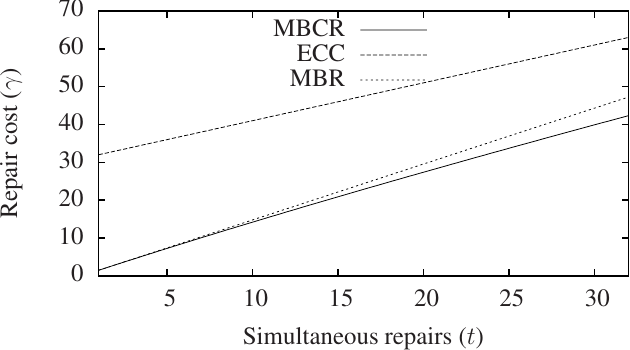}
	\caption{Total repair cost $t\gamma$ for $\MM=32$, $d=48$ and $k=32$. MBCR codes permanently outperform both erasure correcting codes and regenerating codes}
	\label{fig:mbcr}
\end{figure}

Minimum Bandwidth Coordinated Regenerating Codes correspond to optimal codes that provide the lowest possible repair cost (bandwidth consumption) $\gamma$ while minimizing the storage cost $\alpha$.  Figure~\ref{fig:mbcr} compares MBCR codes to  both Dimakis \emph{et al.} 's MBR~\cite{Dimakis2010} and erasure correcting codes with delayed repairs (ECC). 
\begin{align*}%
\alpha=\gamma
&
\beta=&\MB\frac{2}{2d-k+t}%
&
\beta'=&\MB\frac{1}{2d-k+t}%
\end{align*}%

We determine these values in two steps. We study two particular cuts to find  the minimum values required to ensure that the max flow is at least equal to the file size, thus proving the optimality of the solution if correct. We then prove that these quantities are sufficient for all possible cuts.

\begin{IEEEproof}[Proof of MBCR (Optimality)]%
Let us consider two specific successions of repairs ($\mathbf{u}=[1,1,\dots]$ (Fig. \ref{fig:g2}) and $\mathbf{u}=[t,t,\dots]$ (Fig. \ref{fig:g1})). The corresponding repairs are described in the Proof of Theorem \ref{thm:crc}. As we want to minimize $\gamma$ before $\alpha$, we assume $\alpha \ge \gamma$.

When $\forall{}i, u_i=t$, it is required that
\begin{equation*}%
\sum_{i=0}^{\frac{k}{t}-1}t \left( \left(d-\sum_{j=0}^{i-1}{t}\right)\beta\right) \ge \MM %
\end{equation*}%
which is equivalent to
\begin{equation*}%
\beta \ge \MB\frac{2}{2d-k+t} %
\end{equation*}%

When $\forall{}i,  u_i=1$, it is required that
\begin{equation*}%
\sum_{i=0}^{k-1} \left( \left(d-\sum_{j=0}^{i-1}{1}\right)\beta+\left(t-1\right)\beta'\right) \ge \MM %
\end{equation*}%
which is equivalent to
\begin{equation*}%
\beta' \ge \frac{1}{t-1}\left(\MB - \beta\frac{2d-k+1}{2}\right) %
\end{equation*}%

Consider the smallest value $\beta'=\frac{1}{t-1}\left(\MB - \beta\frac{2d-k+1}{2}\right)$, the associated repair cost is $\gamma=\frac{M}{k}+\frac{k-1}{2}\beta$. This implies that the repair cost grows linearly with $\beta$, we therefore seek to minimize $\beta$. The minimum value for $\beta$ is $\MB\frac{2}{2d-k+t}$. 
\end{IEEEproof}%

\begin{IEEEproof}[Proof of MBCR (Correctness)]%
We have proved that the aforementioned values are required for two specific scenarios. We now prove that such values ensure that enough information flows through every cut for any scenario thus proving correctness. According to Theorem~\ref{thm:crc}, the following condition is sufficient for correctness. We show that the values of $\alpha$, $\beta$ and $\beta'$ for MBCR codes satisfy this condition:%
\begin{equation*}%
\sum_{i=0}^{g-1} \left( u_i \min\left\{\left(d-\sum_{j=0}^{i-1}{u_j}\right)\beta+(t-u_i)\beta' , \alpha \right\} \right) \ge \MM %
\end{equation*}%
since $\alpha$ (the stored part) is always larger than or equal to the transmitted data,
\begin{equation*}%
\sum_{i=0}^{g-1}u_i \left( \left(d-\sum_{j=0}^{i-1}{u_j}\right)\beta+(t-u_i)\beta' \right) \ge \MM %
\end{equation*}%
replacing $\alpha$, $\beta$ and $\beta'$ by their values, 
\begin{equation*}%
\sum_{i=0}^{g-1}u_i \left( (d-\sum_{j=0}^{i-1}{u_j})2+(t-u_i) \right)  \ge k(2d-k+t) %
\end{equation*}%
which is equivalent to%
\begin{equation*}%
(2d+t)\sum_{i=0}^{g-1}u_i - \left(\sum_{i=0}^{g-1}u_i\right)^2 \ge k(2d-k+t) %
\end{equation*}%
As $k = \sum_{i=0}^{g-1}{u_i}$, it simplifies to $(2d+t)k- k^2 \ge k(2d-k+t)$
which is always true.  Hence, MBCR codes are correct.%
\end{IEEEproof}%

\subsubsection{MSCR codes}
\label{ssec:mscr}
\begin{figure}[!h]
	\centering
	\includegraphics[width=\xb\linewidth*\real{0.9}]{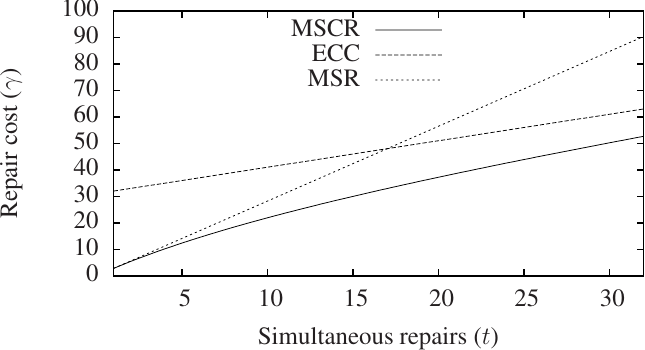}
	\caption{Total repair cost $t\gamma$ for $\MM=32$, $d=48$ and $k=32$. MSCR codes permanently outperform both erasure correcting codes and regenerating codes}
	\label{fig:mscr}
\end{figure}

Minimum Storage Coordinated Regenerating Codes correspond to optimal codes that provide the lowest possible storage cost $\alpha$ while minimizing the repair cost $\gamma$. This point has been independently characterized by Hu \emph{et al.} in~\cite{Hu2010}; however, they assume that $\beta=\beta'$ without proving it. We present a simple derivation from Theorem~\ref{thm:crc} allowing us to characterize this point. Figure~\ref{fig:mscr} compares MSCR codes to  both Dimakis \emph{et al.}'s MSR~\cite{Dimakis2010} and erasure correcting codes with delayed repairs (ECC). Note that for $d=k$, our MSCR codes share the same repair cost as erasure correcting codes delayed repair. Yet, in this case, our codes still have the advantage that they balance the load evenly thus avoiding bottlenecks. 
\begin{align*}%
\alpha&=\MB%
&
\beta=&\MB\frac{1}{d-k+t}%
&
\beta'=&\MB\frac{1}{d-k+t}%
\end{align*}%

\begin{IEEEproof}[Proof of MSCR (Optimality)]%
Let us consider two particular successions of repairs ($\mathbf{u}=[1,1,\dots]$ and $\mathbf{u}=[t,t,\dots]$) leading to the graphs shown on  Figure~\ref{fig:ifg}. The repairs corresponding to such graphs are described in the Proof of Theorem \ref{thm:crc}. 

We minimize $\alpha$ first. It is clear that $\alpha=\MB$ is minimal since $\alpha < \MB$ makes impossible to reconstruct a file of size $\MM$ using only $k$ blocks. Hence, what is important is now that each element of the sum is at least equal to $\MB$. 
\begin{equation*}
\forall{}i\in{0\ldots{}g-1}, (d-\sum_{j=0}^{i-1}{u_i})\beta+(t-u_i)\beta' \ge \MB %
\end{equation*}

When $\forall{}i, u_i=t$ (Fig. \ref{fig:g1}), it is required that
\begin{equation*}%
\forall{}i\in{0\ldots{}k/t-1}, (d-\sum_{j=0}^{i-1}{t})\beta \ge \MB %
\end{equation*}%
which is equivalent to
\begin{equation*}%
\beta \ge \MB\frac{1}{d-k+t} %
\end{equation*}%

When $\forall{}i,  u_i=1$ (Fig. \ref{fig:g2}), it is required that
\begin{equation*}%
\forall{}i\in{0\ldots{}k-1}, (d-\sum_{j=0}^{i-1}{1})\beta+(t-1)\beta' \ge \MB %
\end{equation*}%
which is equivalent to
\begin{equation*}%
\beta' \ge \frac{1}{(t-1)}(\MB - \beta(d-k+1)) %
\end{equation*}%

Consider the smallest value $\beta'=\frac{1}{(t-1)}(\MB - \beta(d-k+1))$, the associated repair cost is  $\gamma=\MB+(k-1)\beta$. This implies that the repair cost grows linearly with $\beta$, we therefore seek to minimize $\beta$. The minimum value for $\beta$ is $\beta=\MB\frac{1}{d-k+t}$.  
\end{IEEEproof}%
 
\begin{IEEEproof}[Proof of MSCR (Correctness)]%
The proof of correctness is quite similar to the previous one. It consists in proving that  
\begin{equation*}
\sum_{i=0}^{g-1} u_i \min\{\alpha,(d-\sum_{j=0}^{i-1}{u_j})\beta+(t-u_i)\beta'\} \ge \MM
\end{equation*}
is always verified when $\alpha$, $\beta$ and $\beta'$ take the aforementioned values. 

Since each element of the sum is at most $u_i\MB$, each element of the sum must satisfy the following constraint.%
\begin{equation*}%
\forall{}i<g,\min\{\MB,(d-\sum_{j=0}^{i-1}{u_j})\beta+(t-u_i)\beta' \} \ge \MB%
\end{equation*}%
Applying values for MSCR codes,%
\begin{equation*}%
\forall{}i<g,\ \frac{1}{d-k+t}(d_i-\sum_{j=0}^{i-1}{u_j}+(t-u_i)) \ge 1%
\end{equation*}%
which is satisfied if %
\begin{equation*}%
\forall{}i <g,\ \sum_{j=0}^{i}{u_j} \le k%
\end{equation*}%
which is true since $\sum_{j=0}^{g-1}{u_j} = k$ and $u_j > 0$. Therefore, MSCR codes are correct.%

\end{IEEEproof}%

%
%
%


\subsubsection{General CR codes}
\begin{figure}[!h]
	\centering
	\includegraphics[width=\xb\linewidth*\real{0.9}]{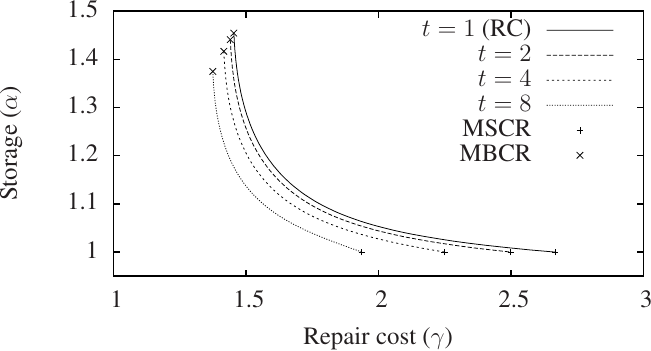}
	\caption{Optimal tradeoffs between storage and repair costs for $k=16$ and $d=24$.  Regenerating codes (RC)~\cite{Dimakis2010} are depicted as $t=1$. For each $t$, both MSCR and MBCR are shown. Costs are normalized by $\MM/k$.}
	\label{fig:tradeoff_t}
\end{figure}

The general case corresponds to all possible trade-offs in between MSCR and MBCR. Valid points $(\alpha, \beta, \beta')$ can be determined by performing a numerical minimization of the repair cost $\gamma$ for various storage cost $\alpha$ under constraints of~\eqref{eq:cons} and~\eqref{eq:obj}. Figure~\ref{fig:tradeoff_t} shows the optimal tradeoffs $(\alpha,\gamma)$: coordinated regenerating codes ($t > 1$) can go beyond the optimal tradeoffs for independent repairs ($t=1$) defined by regenerating codes by Dimakis~\emph{et al.}~\cite{Dimakis2010}.

\section{Adaptive Regenerating Codes}
\label{sec:adapt} 

So far we assumed $t$ and $d$ to remain constant across repairs, similarly to~\cite{Dimakis2010} where $d$ is assumed to remain constant. It may not be realistic in real systems that are dynamic.
%

At the Minimum Storage Point ($\alpha=\MB$), such strong assumptions are not needed as repairs are independent (\textit{i.e.}, each term of the sum in~\eqref{eq:cons} can be treated independently). We propose to adapt the quantities to transfer $\beta$ and $\beta'$ to the system state, which is defined by the number $t$ of devices being repaired and the number $d$ of live devices. The resulting adaptive regenerating codes simplify the system design as only the parameter $k$ needs to be decided during the conception: adaptive regenerating codes decide, at runtime for each repair, the best $(d,t)$  to offer the lowest repair cost $\gamma$.\vspace{-0.05cm}

%
\subsection{Adaptive codes at the Minimum Storage point}
\begin{theorem}
\label{thm:arc}
Adaptive regenerating codes $(k,\Gamma)$ are both correct and optimal. $\Gamma$ is a function $(t,d) \rightarrow (\beta_{t,d},\beta_{t,d}')$ that maps a particular repair setting to the amounts of information to be transferred during a repair.%
\begin{align}%
\beta_{t,d}&=\MB\frac{1}{d-k+t}
&
\beta_{t,d}'&=\MB\frac{1}{d-k+t}
\label{eq:arc}
\end{align}  
\end{theorem}

In this subsection, we prove they are correct and Pareto optimal.%
\begin{lemma}%
For any information flow graph $\MG$ compounded of initial devices that obtain $\alpha$ bits directly from the source $\mathit{S}$ and of additional devices that join the graph in groups of $t_i$ devices obtaining $\beta_{t_i,d_i}$ from  $d_i$ existing devices and $\beta'_{t_i,d_i}$ from each of the other $t_i-1$ joining devices, any data collector $\mathit{DC}$ that connects to a subset of $k$ out-nodes  of $\MG$ satisfies:
\begin{gather}%
\mathrm{mincut}(\mathit{S},\mathit{DC}) \ge  \\\notag
\min_{\mathbf{u} \in P}\left(\sum_{i=0}^{g-1} u_i \min\{\alpha,(d_i-\sum_{j=0}^{i-1}{u_j})\beta_{t_i,d_i}+(t_i-u_i)\beta'_{t_i,d_i}\}\right) 
\label{eq:adaptcons}
\end{gather}
with $P = \{\mathbf{u} : 1 \le u_i \le t_i \land \sum_{i=0}^{g-1}{u_i}=k \}$.
\label{lem:globadapt}
\end{lemma}

\begin{IEEEproof}%
The proof is similar to the proof of Lemma~\ref{lem:glob}.%
\end{IEEEproof}%
\begin{IEEEproof}[Proof of Theorem~\ref{thm:arc} (Correctness)]%
Using Lemmas~\ref{lem:info},~\ref{lem:info2} and~\ref{lem:globadapt}, we can define the following sufficient condition for the code to be correct. The condition is satisfied when $\beta$ and $\beta'$ take the values defined in~\eqref{eq:arc}.%
\begin{gather}%
 \forall \mathbf{u}, \textrm{ such that } \sum_{i=0}^{g-1}{u_i}=k  \textrm{ and } 1 \le u_i \le t, \notag \\
\sum_{i=0}^{g-1} u_i \min\{\MB,(d_i-\sum_{j=0}^{i-1}{u_j})\beta_{t_i,d_i}+(t_i-u_i)\beta'_{t_i,d_i} \} \ge \MM \notag %
\end{gather}%
The condition must be satisfied for every $\mathbf{u}$. For any $\mathbf{u}$, since each element of the sum is at most $u_i\MB$, each element of the sum must satisfy the following constraint.%
\begin{equation*}%
\forall{}i<g,\min\{\MB,(d_i-\sum_{j=0}^{i-1}{u_j})\beta_{t_i,d_i}+(t_i-u_i)\beta'_{t_i,d_i} \} \ge \MB%
\end{equation*}%
Applying formulas of~\eqref{eq:arc},%
\begin{equation*}%
\forall{}i<g,\ \frac{1}{d_i-k+t_i}(d_i-\sum_{j=0}^{i-1}{u_j}+(t_i-u_i)) \ge 1%
\end{equation*}%
which is satisfied if %
\begin{equation*}%
\forall{}i \le g-1,\ \sum_{j=0}^{i}{u_j} \le k%
\end{equation*}%
which is true since $\sum_{j=0}^{g-1}{u_j} = k$ and $u_j > 0$. Therefore, adaptive regenerating codes are correct.%
\end{IEEEproof}%
\begin{IEEEproof}[Proof of Theorem~\ref{thm:arc} (Optimality)]%
We prove by contradiction that the adaptive regenerating codes are optimal. Let us assume that there exists a correct code $(k,\bar{\Gamma})$ such that $\bar{\Gamma}<\Gamma$ (i.e., for some $(t,d)$, $\bar{\Gamma}(t,d) < \Gamma(t,d)$).

Consider a set of failures such that all repairs are performed by groups of $t$ devices downloading data from $d$ devices. Consider the corresponding information flow graph. Assuming repairs are performed with a correct code $(k,\bar{\Gamma})$, the information flow graph also corresponds to a correct code $(d+t,k,t,d,\alpha,\bar{\beta}_{t,d},\bar{\beta}'_{t,d})$. 

Moreover, according to the previous section, these failures can be repaired optimally using the MSCR code $(d+t,k,t,d,\alpha,\beta_{t,d},\beta'_{t,d})$. Therefore, there is a contradiction since the code $(d+t,k,t,d,\alpha,\bar{\beta}_{t,d}, \bar{\beta}'_{t,d})$ cannot be correct if the code $(d+t,k,t,d,\alpha,\beta_{t,d},\beta'_{t,d})$ is optimal. A correct code $(k,\bar{\Gamma})$ cannot exist, and the adaptive regenerating code $(k,\Gamma)$ defined in this section is optimal.
\end{IEEEproof}
\vspace{1eX}

Building on results from coordinated regenerating codes (especially MSCR), we have defined adaptive regenerating codes and proved that they are both correct and optimal. These codes are of particular interest for dynamic systems where failures may occur randomly and simultaneously.

\changed
\subsection{Adaptive codes at the Minimum Bandwidth point}
We have built adaptive regenerating codes from Minimum Storage codes ($\alpha=\MB$) by observing that initial assumptions of fixed value $d$ and $t$ can be relaxed.  In this subsection, we study whether adaptive codes can be built from MBR codes. We determine lower bounds on the storage and repair cost $\gamma_{d,t}$. These lower bounds allow concluding that an adaptive scheme at the Minimum Bandwidth point cost as much as classical erasure correcting codes.

Let us consider that $d_i$ can take any value between $d_{\mathrm{min}}=k$ and $d_{\mathrm{max}}=n-1$, and that  $t_i$ can take any value between $t_{\mathrm{min}}=1$ and $t_{\mathrm{max}}=k$. Since we cannot predict the future, when choosing $\beta_i$ and $\beta'_i$, we must assume any value for $d_j$ and $t_j$ with $j > i$. More specifically, the current repair can be the first of a sequence since we do not know which devices will fail, how they will be repaired and how data will be collected. We need to consider the worst case that can occur in the future: $d_i=d_{\mathrm{min}}$ and $t_j=t_{\mathrm{min}}$. 

In a first scenario when $\forall{}i, u_i=t_i$ (\emph{e.g.}, Fig. \ref{fig:g1}), it is required that
\begin{equation*}%
\sum_{i=0}^{g-1}t_i \left( (d_i-\sum_{j=0}^{i-1}{t_i})\beta_{d_i,t_i}\right) \ge \MM %
\end{equation*}%
which expands to
 \begin{align*}%
 (t_0-1)d_0\beta_{t_0,d_0}- \sum_{i=1}^{g-1}t_0\beta_{t_i,d_i}+ \sum_{i=1}^{g-1}t_i \left( (d_i-\sum_{j=1}^{i-1}{t_i})\beta_{t_i,d_i}\right) &
 \\ \ge \MM %
\end{align*}%
replacing $d_i$ and $t_i$ by their minimum admissible values since we cannot make any assumption on the future, and $g$ by the number of groups when all groups but the first are of size $t_i=1$ (\emph{i.e.}, $g=k-t_0+1$), we get
 \begin{align*}%
(t_0-1)d_0\beta_{t_0,d_0}-  \sum_{i=1}^{k-t_0}t_0\beta_{1,k}+
\sum_{i=1}^{k-t_0}\left( k-\sum_{j=1}^{i-1}{1}\right)\beta_{1,k} \ge \MM %
\end{align*}%
which simplifies to
 \begin{equation*}%
(t_0-1)d_0\beta_{t_0,d_0}-(k-t_0)\beta_{1,k}+\sum_{i=1}^{k-t_0} \left( (k-i+1)\beta_{1,k}\right) \ge \MM %
\end{equation*}%
which simplifies to
 \begin{equation*}%
(t_0-1)d_0\beta_{t_0,d_0}+\sum_{i=1}^{k-t_0} k\beta_{1,k} - \sum_{i=1}^{k-t_0}i\beta_{1,k} \ge \MM %
\end{equation*}%

Let us consider the case of $t_0=1$ and $d_0=k$, and determine the minimal value for $\beta_{1,k}$.
 \begin{equation*}%
(1-1)k\beta_{k,1}+(k-1)k\beta_{1,k}-\frac{(k-1)(k-1+1)}{2}\beta_{1,k}\ge \MM %
\end{equation*}%
 \begin{equation*}%
\beta_{1,k} \ge \MM \frac{2}{k-1}%
\end{equation*}%

The first scenario we considered allowed determining that $\beta_{1,k}>\MM\frac{2}{k-1}$. We now consider another possible scenario to obtain a lower bound on the per failed node repair cost $\gamma_{t_0,d_0}$.  In this second scenario, when $\forall{}i,  u_i=1$ (\emph{e.g.}, Fig. \ref{fig:g2}), it is required that
\begin{equation*}%
\sum_{i=0}^{k-1} \left( (d_i-\sum_{j=0}^{i-1}{1})\beta_{d_i,t_i}+(t_i-1)\beta'_{d_i,t_i}\right) \ge \MM %
\end{equation*}%
which expands to
\begin{align*}%
\gamma_{t_0,d_0}+
\sum_{i=1}^{k-1} \left( (d_i-\sum_{j=0}^{i-1}{1})\beta_{t_i,d_i}+(t_i-1)\beta'_{t_i,d_i}\right) \ge \MM %
\end{align*}%
replacing $d_i$ and $t_i$ by their minimum admissible values ($d_i=k$ and $t_i=1$)  since we cannot make any assumption on the future, and $g$ by the number of groups when all groups but the first are of size $u_i=1$ (\emph{i.e.}, $g=k$), we get
\begin{equation*}%
\gamma_{t_0,d_0}+\sum_{i=1}^{k-1} \left( (k-i)\beta_{1,k}\right) \ge \MM %
\end{equation*}%
which simplifies to
\begin{equation*}%
\gamma_{t_0,d_0}+\frac{k(k-1)}{2}\beta_{1,k} \ge \MM %
\end{equation*}%
Hence, we obtain the following lower bound for an adaptive regenerating code operating at the MBR point.
\begin{equation*}%
\gamma_{t_0,d_0} \ge \MM\frac{2}{k+1} %
\end{equation*}%
The cost of a static scheme assuming that we contact as few nodes as possible $d=k$ and repair as few nodes as possible $t=1$ is $\bar{\gamma}_{k,1}=\MM\frac{2}{k+1}$ as explained in Section~\ref{ssec:mbcr}. Hence, 
\begin{equation*}%
\gamma_{t_0,d_0} \ge \bar{\gamma}_{1,k} %
\end{equation*}%

As a consequence, an adaptive scheme at the Minimum Bandwidth point is meaningless since it would be more expensive than a simpler static MBCR code set up for the worst case (\emph{i.e.}, $t=1$ and $d=k$).

\color{black}

\subsection{Performance}
We compare our Adaptive Regenerating Codes at the MSR point to MFR codes defined in~\cite{Wang2010}. This approach is built upon MSR codes defined by Dimakis \emph{et al.} in~\cite{Dimakis2010}. The coding scheme can be described  as $(k,\Gamma')$ where $\Gamma'$ is a function $d \rightarrow \beta_d$. The $t$ repairs needed are performed independently.
\begin{equation}
\beta_{d}=\MB\frac{1}{d-k+1}
\label{eq:arc0}
\end{equation}
\endchanged

\begin{figure}[!h]
	\centering
	\includegraphics[width=\xb\linewidth*\real{0.9}]{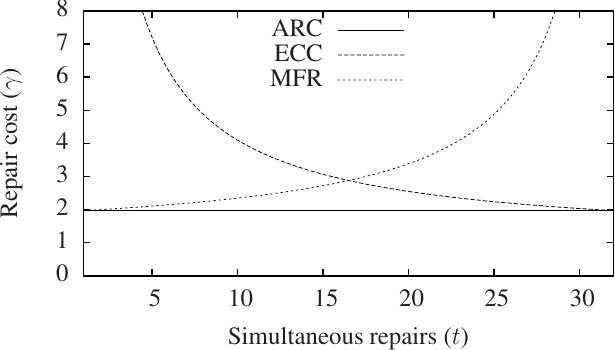}
	\caption{Average repair cost $\gamma$ for $n=64$ and $k=32$. Adaptive Regenerating Codes (ARC) permanently outperform both erasure correcting codes (ECC) and the MFR codes.}
	\label{fig:arc}
\end{figure}

Let us consider the particular case where $d+t=n$. The average cost per repair of our codes remains constant $\gamma=\MB\frac{n-1}{n-k}$. In the MFR approach, which requires repairs to be performed independently,  the average repair cost $\gamma'=\MB\frac{n-t}{n-t-k+1}$ increases with $t$. Therefore, the performance of our adaptive regenerating codes does not degrade as the number of failures increases, as opposed to the MFR constructed upon Dimakis \emph{et al.} 's codes. This is also shown on Figure~\ref{fig:arc}.

\changed

\subsection{Adaptive Coding Schemes}
\label{sec:arcimpl}
Our approach also has significant advantages over the MFR approach with respect to the actual coding scheme being implemented when $d+t$ is constant. The coding schemes are similar in principle to the one described in Subsection~\ref{ssec:repair} and Figure~\ref{fig:implem}. The only difference is that the values $d$ and $t$ may differ from one repair to the other. Each device stores sub-blocks of data and combines them to send the appropriate quantities of information. To be able to send $\beta=\frac{2}{3} \MB$, each device must store $z=3$ sub-blocks. To be able to send $\beta=\frac{1}{3} \MB = \frac{4}{12}\MB$ or $\beta=\frac{1}{4}\MB=\frac{3}{12}\MB$, each device must store $z=\lcm{\{3,4\}}=12$ sub-blocks.  Hence, the length of any random linear code used to implement such a system is $l=zk$ where $z$ is the number of sub-blocks stored by each device. We now consider a system of constant size $n=d+t$ and compare both implementations.

The implementation of the MFR approach implies that to support $d \in \{k\dots{}n-1\}$, each device must be able to send all quantities $\beta \in \{\frac{1}{1}\MB\dots\frac{1}{n-k}\MB\}$. Hence, $z=\lcm{\{1\dots{}n-k\}}$. It is known that $2^{n-k} \le z \le 3^{n-k}$. Hence, the length of the codes required to implement such codes grows exponentially with $n-k$.

The implementation of our approach implies that to support $d \in \{k\dots{}n-1\}$ as long as $n=d+t$ (\emph{i.e.}, all devices are either alive or being repaired), each device must be able to send quantities $\beta=\frac{1}{n-k}\MB$ and $\beta'=\frac{1}{n-k}\MB$. Hence, $z=n-k$.  Hence, the length of the codes required to implement such codes grows linearly with $n-k$, and is much smaller than for MFR codes. This is very important since $z$ has a direct impact on the computational complexity of all operations (encoding, recoding, and decoding).

\section{Lazy repairs in regenerating codes}
\label{sec:lazy}
By supporting the repair of multiple failures, we enable delayed or lazy repairs. They consist in deliberately delaying repairs so that multiple repairs are performed simultaneously thus factoring some costs. Lazy repairs have successfully been applied to regular erasure correcting codes so as to reduce network repair costs~\cite{Bhagwan2004,Datta2006,Dalle2009}. We study the impact on network-related and disk-related costs of lazy repairs applied to coordinated regenerating codes.

\subsection{Network repair cost}
As previously explained, in regenerating codes, the higher the number of devices being contacted $d$, the higher the savings on the repair cost $\gamma$. Moreover, when repairs are delayed, higher values for the number of devices being repaired $t$ lead to higher savings on the repair cost $\gamma$. If we consider a system of constant size $n=d+t$, these two objectives are contradictory: the longer the delay, the lower the number of live devices $d$. An interesting question is what is the optimal threshold  $t$ for triggering repairs assuming that $d+t$ is constant (i.e., is it useful to deliberately delay repairs?). This question is addressed hereafter by studying how MBCR codes and MSCR codes behave as $t$ changes in a system of constant size.

\begin{theorem}
If we consider a system of size $n=d+t$, for MBCR codes, the optimal value is $t=1$ while for MSCR codes any value $t \in \{1\dots{}n-k\}$ is optimal.
\end{theorem}
\begin{IEEEproof}
Let us consider the repair cost assuming that $n=d+t$ is constant. For MBCR codes, the cost $\gamma=\MB\frac{2n-t-1}{2n-k-t}$ increases when $t$ increases. The optimal value of $t$ for MBCR codes is the lowest possible value (i.e., $t=1$).  For MSCR codes, the cost $\gamma=\MB\frac{n-1}{n-k}$ does not depend on $t$. The repair cost of MSCR remains constant, and $t$ can be set to any value as there is no optimum. Neither MSCR nor MBCR allow additional gains by deliberately delaying repairs (i.e., deliberately setting $t>1$).  
\end{IEEEproof}

\begin{corollary}
If we consider a system of size $n=d+t$ where $t$ can be freely chosen (i.e., the value of $t$ is not constrained by the system)  both MSR and MBR regenerating codes~\cite{Dimakis2010} are optimal. Hence deliberately delaying repairs to force high values for $t$ does not bring additional savings. 
\end{corollary}

\changed
\subsection{I/O and disk reads}
\label{lr-io}
In spite of the absence of improvement with regard to network-related repair cost, lazy repairs in coordinated regenerating codes can reduce disk-related costs. The impact of repairs on disks can be measured by two metrics: \emph{(i)} the number of disk accesses (\emph{i.e.}, number of disk I/O), and \emph{(ii)} the amount of data read on disks (\emph{i.e.}, disk bandwidth) which is designated as access in~\cite{Tamo2012}. Minimizing these metrics is of interest since the repair process should put as little pressure as possible on non-failed devices to limit the impact of unavailability and failures on the rest of the system.

\subsubsection{Number of accessed disks (I/O)}
The number of accessed disks during a repair is $d$. Each disk accessed must be woken up and perform one I/O operation. If $t$ repairs are performed independently, each repair implies $d$ accesses leading to a total of $td$ accesses. However, if multiple repairs are performed in a coordinated way, the $t$ repairs imply only $d$ access.

\begin{figure}[!h]
	\centering
	\includegraphics[width=\xb\linewidth*\real{0.9}]{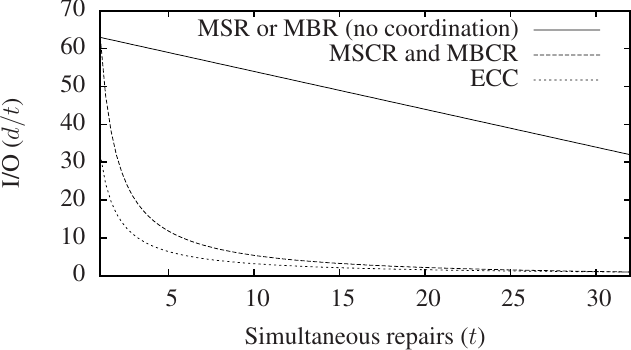}
	\caption{Average I/O $d$ for $n=64$ and $k=32$. Lazy repair in coordinated regenerating codes (MSCR or MBCR) significantly reduces the I/O costs, similarly to what is observed in regular erasure correcting codes (ECC) and contrary to what is observed with regenerating codes (MSR or MBR). }
	\label{fig:io_io}
\end{figure}

Let us consider a system of constant size ($n=d+t)$. As explained previously, in this case, MSCR codes with $d=n-t$ have the same network repair cost as MSR codes with $d=n-1$. Furthermore, delayed repair imply less live devices $d$ involved leading to an even lower number of disk accesses. Instead of accessing $t(n-1)$ live devices for performing $t$ successive repairs, coordinated regenerating codes allow accessing only $n-t$ live devices to perform $t$ simultaneous repairs. Hence, even low values of $t$ significantly reduce the impact on live disks in term of I/O, as shown on Figure~\ref{fig:io_io}, which plots the disk accessed per device repaired.

\subsubsection{Amount of data read on disks (disk bandwidth)}
We now consider the amount of data read on disks which has an impact on the disk bandwidth. When using regenerating codes not specifically optimized for reducing the impact on disks (\emph{e.g.}, randomized codes~\cite{Dimakis2010}, or many exact codes~\cite{Rashmi2011,Shah2012,Suh2011}), each of the $d$ devices contacted read all data they store ($\alpha$) and compute some $\beta$ linear combinations of this data. For simultaneous repairs with coordinated regenerating codes, each of the $d$ devices contacted reads all the data they store and compute $t\beta$ linear combinations of this data. In both case (for one independent repair, or for $t$ simultaneous repairs) the amount of data read is $d\alpha$. Moreover, if we consider a system of constant size ($n=d+t$) (and thus MSCR codes with a network repair cost identical to MSR codes), coordinated regenerating codes imply reading $(n-t)\alpha$ for $t$ coordinated simultaneous repairs.

\begin{figure*}[t]
\centering
\subfloat[Functional repair]{\centering
        \includegraphics[height=0.17\linewidth,trim=0 -35 0 0]{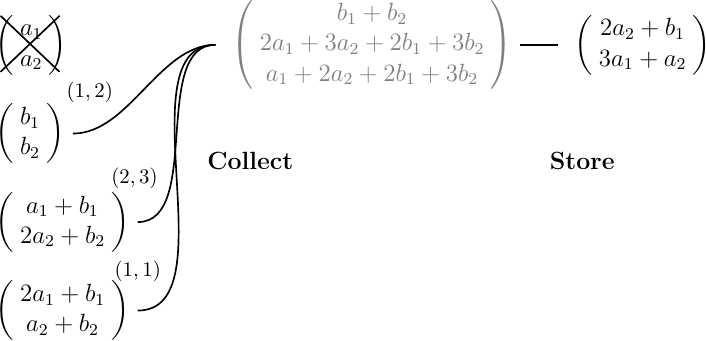}
        \label{fig:functional}
} \hfil
\subfloat[Exact repair (scalar $\beta=1$)]{\centering
        \includegraphics[height=0.17\linewidth,trim=0 -35 0 0]{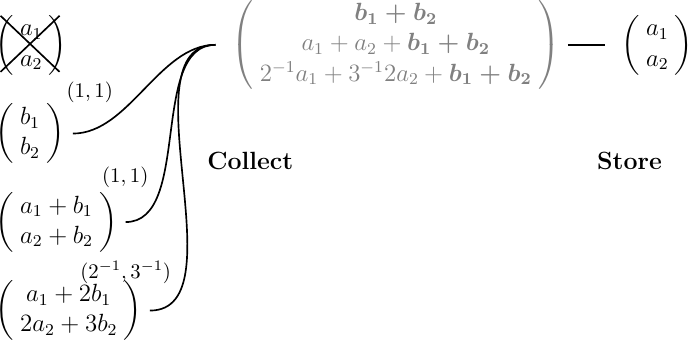}
        \label{fig:exact}
}\hfil
\subfloat[Exact repair (vector $\beta=2$)]{\centering
        \includegraphics[height=0.18\linewidth]{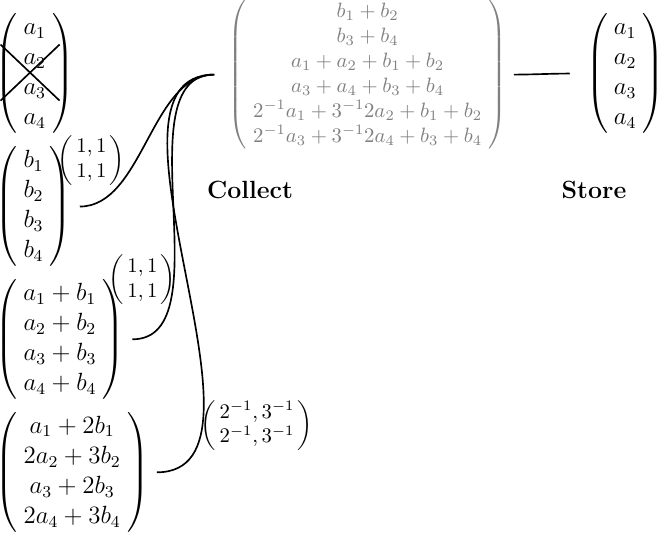}
        \label{fig:sub}
}
\caption{Regenerating codes can be repaired functionally or exactly. In our example, the device storing $(a_1,a_2)$ fails and is regenerated. When relying on functional repairs, the information about $(a_1,a_2)$ is regenerated but not in the same form, while when relying on exact repairs, $(a_1,a_2)$ is regenerated exactly. This figure also illustrates the difference between scalar codes where scalar are transmitted over the network and vector codes where vectors are sent over the network.}
\label{fig:soa}
\end{figure*}

\begin{figure}[!h]
	\centering
	\includegraphics[width=\xb\linewidth*\real{0.9}]{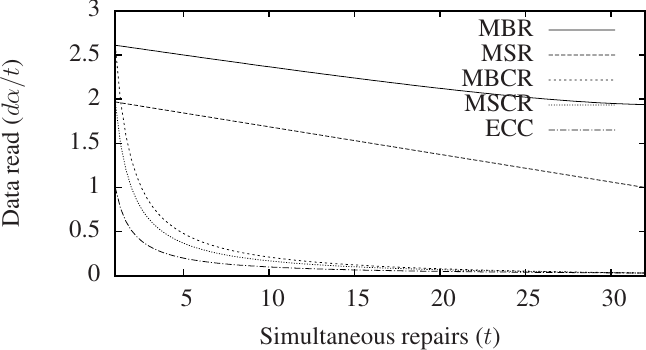}
	\caption{Average amounts of data read $d\alpha$ for $n=64$ and $k=32$. Lazy repair in coordinated regenerating codes (MSCR or MBCR) significantly reduces the disk bandwidth consumption, similarly to what is observed in regular erasure correcting codes (ECC) and contrary to what is observed with regenerating codes (MSR or MBR).}
	\label{fig:io_rd}
\end{figure}

Figure~\ref{fig:io_rd} plots the amount of data read when delaying repairs. When repairs are delayed (\emph{i.e.}, lazy repairs), the number of contacted disks $d$ as well as the total amount of data read on disks are both reduced by approximately a factor $t$. Furthermore, the network repair cost is kept constant or only slightly increased when delaying repairs. Hence, lazy repairs are interesting when the impact on non-failed disks must be limited.

The bounds on the amount of data read in regenerating codes are not tight (contrary to the bounds on the amounts of data transferred). More specifically by carefully building exact regenerating codes for single failures, it has been shown that the amount of data read on disk can be reduced either at the price of computational complexity~\cite{Cadambe2011b,Tamo2011,Tamo2012}, or at the price of a reduced storage efficiency~\cite{Rashmi2009,ElRouayheb2010}. In this section, we have shown that using existing and simple constructions (\emph{e.g.,} randomized codes), lazy repairs can reduce the I/O costs and the amount of data read on disks.  Hence, an interesting perspective would be to determine tight achievable bounds for disks I/O costs for both coordinated and regular regenerating codes. Indeed, since coordinated regenerating codes allow lowering the impact on non-failed devices for randomized codes, it may be interesting to study lazy repairs as a way to reduce the disk I/O costs for exact regenerating codes as it may allow further savings when compared to codes supporting only single repairs~\cite{Cadambe2011b,Tamo2011,Tamo2012}.

\endchanged

\section{Related Work}
\label{sec:rel}

\subsection{Exact Regenerating Codes}

Regenerating codes (including coordinated regenerating codes) can be implemented using random linear codes~\cite{Li2003,Koetter2003,Ho2006}. In this case, repairs are termed as functional repair (Figure~\ref{fig:functional}) for the regenerated data is not strictly equal to the lost data. However, such non-deterministic schemes are not desirable for they \emph{(i)} require homomorphic hash functions to provide basic security (integrity checking), \emph{(ii)} cannot be turned into systematic codes, which offer access to data without decoding, and \emph{(iii)} can only provide probabilistic guarantees. Deterministic schemes overcome these issues by offering exact repair (\emph{i.e.}, during a repair, the regenerated block is equal to the lost block and not only \emph{equivalent} as shown on Figure ~\ref{fig:exact}). Yet, it has been shown that exact repair is strictly harder than functional repair~\cite{Shah2010b,Shah2012} , which means that the existence of functional regenerating codes does not imply that exact regenerating codes exist. Hence, an interesting question is whether the previous tradeoffs, which apply to functional repairs, can still be achieved for exact repairs.

\begin{figure*}[t]%
\includegraphics[width=\linewidth]{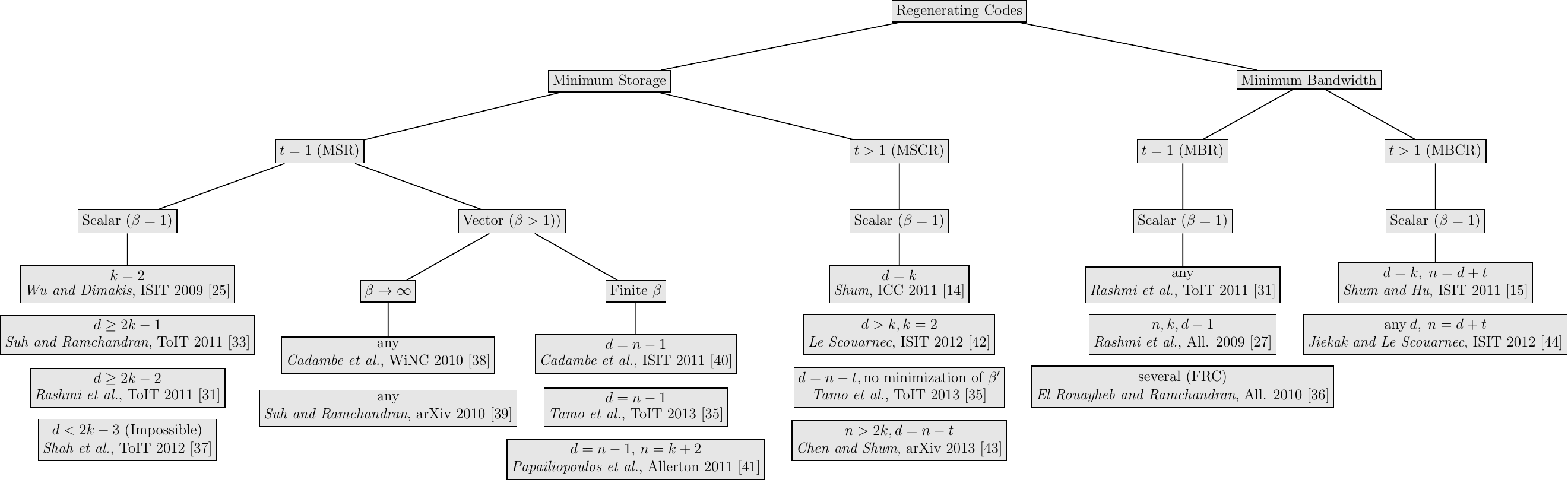}%
\caption{A partial taxonomy of results on exact regenerating codes.}%
\label{fig:soatree}%
\end{figure*}

Figure~\ref{fig:soatree} gives an overview of results related to the construction of exact regenerating codes. Two main classes of codes exist, namely scalar and vector codes.  Scalar codes rely on indivisible sub-blocks of size $\beta=1$ as shown on Figure~\ref{fig:exact}. Yet, scalar codes are not always sufficient as explained hereafter. Hence vector codes, relying on sub-packetization, have been defined. In these codes, manipulated sub-blocks are smaller than the smallest amount of information to be transmitted (\emph{i.e.}, sub-blocks are of size $\frac{\beta}{r}$ such that to $r$ indivisible sub-blocks are transmitted when sending $\beta=r$) as shown on Figure~\ref{fig:sub} where $\beta=2$.

Among all possible regenerating codes, most of the studies have focused on the minimum storage point. For MSR codes that are able to repair single failures ($t=1$), studies have heavily relied on interference alignment, first applied to $k=2$ in \cite{Wu2009}. The best known scalar codes either use interference alignment~\cite{Suh2011} to allow  $d \ge 2k-1$, or use the product matrix framework~\cite{Rashmi2011} to allow $d \ge 2k-2$.  However, scalar codes cannot be used to achieve $d <  2k-3$ as shown in~\cite{Shah2010b}. 

To circumvent this impossibility of constructing scalar MSR codes when $d<2k-3$, vector codes (\emph{i.e.}, $\beta>1$) have been proposed. Vector codes supporting exact repair can be built for any values $n,k,d$ when $\beta  \rightarrow \infty$~\cite{Cadambe2010,Suh2010a}. However, these constructions require infinite sub-packetization and, hence, are not practical.  Recent works~\cite{Cadambe2011, Tamo2011} have shown that finite sub-packetization $\beta=(n-k)^k$  is sufficient to perform exact repair of the systematic devices leading to practical codes. The repair of all devices is possible when $d=n-1, n=k+2$ as shown in~\cite{Papailiopoulos2011}. As a result, the exact repair of all devices with vector MSR codes is not fully solved.

For the case of multiple failures $t>1$, only scalar MSCR codes ($\beta=1$) have been considered. Initially, \cite{Shum2011} considered the degenerated case of $d=k$  where the repair boils down to repairing  in parallel $t$  independent erasure correcting codes. Later, interference alignment has been used to build exact MSCR codes for $k=2, d \ge k$~\cite{LeScouarnec2012a}, or to enable the repair of multiple failures in exact MSR codes defined by Suh and Ramachandran~\cite{Chen2013}. Exact codes by Tamo~\emph{et al.}~\cite{Tamo2011} also support the repair of multiple failures but do not minimize the network traffic between devices being repaired (\emph{i.e.,} $(t-1)\beta'$).
 
With respect to the MBR point, the best known construction~\cite{Rashmi2011} are scalar codes based on the product matrix framework and allow the repair for any value of $n,k,d$. Some interesting alternative codes~\cite{Rashmi2009,ElRouayheb2010} allow repair by transfer (\emph{i.e.}, without performing any linear operation) and rely on fractional repetition codes. 

For the repair of multiple failures in MBCR codes, Shum \emph{et al.} again consider the case of $d=k$ and map to repairing $t$ independent erasure correcting codes~\cite{Shum2011b}. Jiekak \emph{et al.} have designed a scheme~\cite{Jiekak2012} that is not restricted to $d=k$ and works for any value $k$, $d$ and $t$ as long as $n=d+t$.

Finally, regenerating codes~\cite{Dimakis2007,Dimakis2010} can be extended into adaptive regenerating codes. The first supports repairing multiple failures optimally and has a constant $\beta$ as long as $n=d+t$ (\emph{i.e.}, as long as the total system size $n$ including both live devices and failed devices being repaired remains constant) thus making implementation easier (Section~\ref{sec:arcimpl}). The existence (resp. non-existence) of exact adaptive regenerating codes is strongly tied to the existence (resp. non-existence) of exact MSCR codes. The two known constructions~\cite{LeScouarnec2012a,Chen2013} of MSCR codes can be used to implement adaptive regenerating codes where $d+t$ is a constant.

\subsection{Variations on Regenerating Codes}
Regenerating codes~\cite{Dimakis2010} and coordinated regenerating codes assume a symmetric role for all devices (\emph{i.e.}, they all transfer the same amounts of information). Since network connections between every device may not be equivalent, it is interesting to adapt the repair strategy to take into account the underlying network topology. A first study~\cite{Li2010} has focused on  structuring the repair as a tree instead of a star. Instead of receiving data directly from live devices, failed devices may receive indirectly data through other failed devices. This can avoid a potential bottleneck links in some specific networks where devices cannot contact all other devices directly (\emph{e.g.}, when devices are connected in a mesh network). Another study~\cite{Shah2010} has focused on downloading unequal amounts of information from other devices during repairs. They define the total amount of information that must be downloaded depending on the maximum amount of information that can be downloaded from each device. They show that the lowest repair cost is offered when all devices download the same amount of data (i.e., regular regenerating codes). It is simple to apply the methodology of this last study to our codes and thus to show that allowing unequal downloads (\emph{i.e.}, a non-symmetric system) increases the global repair cost.

Independently from our result, the work~\cite{Hu2010} addresses a subset of the problem we consider. They notice that regenerating codes can only repair single failures and come up with a solution that can handle multiple failures. They naturally define a similar repair method (i.e., they add a coordination step to the information flow graph). Yet, their solution is more limited than ours as they only study the Minimum Storage case (MSR).  Not only, we also study the Minimum Bandwidth (MBR) point, but this  cannot be covered by their model since they assume all transfers are equal (i.e., $\beta=\beta'$). Finally, we also determine numerically the general case (i.e., points between Minimum Storage (MSR) and Minimum Bandwidth (MBR) points). Their paper is also restrictive with respect to system they consider as, they assume a system of constant size where all devices are involved (i.e., $n=d+t$) and do not prove that $\beta=\beta'$ for the MSR point. Finally, we do build upon our result to define an adaptive form of regenerating codes that is more flexible to use in practical systems while they do not consider such constructions.  Hence, the previously published paper~\cite{Hu2010}, which is yet another proof of the importance of the considered problem, covers only a subset of our results even if it shares both the problem and some tools used (an adaptation of Information Flow Graphs from Dimakis et \emph{al.}~\cite{Dimakis2007,Dimakis2010}) .

\subsection{Locally Repairable Codes}
Regenerating codes reduce the repair cost by contacting more devices ($d>k$) but downloading less ($\beta << \alpha$) data from each. An alternative to reduce the repair cost is to contact less devices ($r < k$) while downloading all (or most) data ($\alpha$) from each. Such codes~\cite{Duminuco2008,Oggier2010,Papailiopoulos2012a,Papailiopoulos2012,Oggier2011,Huang2007,Huang2012,Huang2008,Prakash2012} are locally repairable and have been studied for they reduce both network-related cost and disk-related (I/O and disk bandwidth). They work by ensuring that every encoded block can be recomputed from only a small specific subset of other $r$ encoded blocks. Even though this approach offers a reasonably low repair cost, they are not optimal with respect to the repair trade-off as they cannot outperform a regenerating code with $d=n-1$. However, the fact that they reduce disk-related costs is an appealing property for practical deployments. 

\section{Conclusion}
\label{sec:conc}
We proposed \emph{coordinated regenerating codes} supporting simultaneous repairs in regenerating codes. Such codes outperform regenerating codes~\cite{Dimakis2010} when multiple failures are detected and repaired simultaneously. We also proposed \emph{adaptive regenerating codes} that allow adapting the repair strategy to the current state of the system so that it always performs repairs optimally. Based on these codes, we have studied the impact of lazy repairs (\emph{i.e.}, delayed repairs) on regenerating codes: we have shown that while lazy repairs cannot help to reduce network-related repair costs, they can help to reduce disk-related repair costs. 

We focused on functional repair with optimal repair bandwidth. A first perspective is to define exact coordinated or adaptive regenerating codes, as done in~\cite{Shum2011,Shum2011b,LeScouarnec2012a,Jiekak2012,Chen2013}. A second perspective is to use coordinated regenerating codes with lazy repairs as a way to reduce the disk-related costs (I/O and disk bandwidth) in regenerating codes and to define exact coordinated regenerating codes that are optimal with respect to this (similarly to optimal access codes for single failures~\cite{Tamo2012}).

\begin{figure}[h]%
\centering
\small
\def\allsym{\star}
\begin{tabular}{|l|c|c|c|c|c|}%
\hline%
                                        		 & $\bm{d}$ & $\bm{r}$ & $\bm{t}$ \\\hline%
Erasure codes 	               		         & $k$     &  $k$   & $1$    \\\hline%
Erasure codes  (lazy repairs)    	         & $k$     &  $k$   & $\allsym$     \\\hline%
\textbf{Coordinated regenerating codes}		 & $\allsym$     &  $d$   & $\bm{\allsym}$     \\\hline%
Regenerating codes		    	         & $\allsym$     &  $d$   & $1$    \\\hline%
\textbf{Locally repairable regenerating codes}   & $\bm{\allsym}$     &  $\allsym$   & $1$     \\\hline%
Locally repairable codes		        & $n-1$     &  $\allsym$   & $1$     \\\hline%
\end{tabular}%
\caption{Coordinated regenerating codes combine lazy repairs with regenerating codes (stars indicate that the codes do not restrict the value of the corresponding parameter). Similarly, an interesting perspective would be to define locally repairable regenerating codes that would combine regenerating codes with locally repairable codes. Note that the table uses $d$ with its meaning for regenerating codes (i.e., the subset of \emph{any} $d$ nodes that are available for the repair): this differs from its use in some papers where it designated distance or repair degree in locally repairable codes. Indeed, regenerating codes accept \emph{any} $d$ nodes for the repair, while locally repairable codes require \emph{some specific} $r$ nodes among the $d=n-1$ nodes other than the node repaired.\vspace{-1eX}}%
\label{tab:conc}%
\end{figure}%

We intended at studying how regenerating codes and lazy repairs can be combined. The \emph{coordinated regenerating codes} that we propose can be viewed as a  global class of codes that encompass erasure correcting codes with lazy repairs ($d=k$), regenerating codes ($t=1$), erasure correcting codes ($t=1$ and $d=k$), and new codes ($t>1$ and $d>k$) that combine, previously incompatible, existing approaches of regenerating codes and lazy repairs. Similarly, an interesting perspective would be to combine (coordinated) regenerating codes with locally repairable codes so as to be able to compare them and evaluate if a combination can bring improvement. On one side, regenerating codes fetches data from any $d$ available devices among $n-1$ devices. On the other side, locally repairable codes fetches data from $r$ chosen devices among $n-1$ available devices. Locally repairable regenerating codes would fetch data from $r$ chosen devices among any $d$ available devices among $n-1$ devices. Such a model would encompass regenerating codes ($r=d$), locally repairable codes ($d=n-1$) and erasure correcting codes ($r=k, d=k$), as well as a new class of codes ($r<d, d<n-1$). Some existing codes (\emph{e.g.,}~\cite{Oggier2011}),  which support multiple alternatives for local repair could belong to this new class of codes; yet it is not known if they are optimal. The interest of multiple alternatives for local repair is that only $d$ among the $n-1$ devices may be available for repair thus limiting the possible choices for the $r$ devices from which to download data.

\section*{Acknowledgment}
This study was partially funded by the ODISEA (Open Distributed Networked Storage Architecture) collaborative project from the competitiveness clusters System@tic and Images \& Réseaux.

\bibliographystyle{IEEEtran}
\bibliography{regenerating}

\begin{thebibliography}{10}
\providecommand{\url}[1]{#1}
\csname url@samestyle\endcsname
\providecommand{\newblock}{\relax}
\providecommand{\bibinfo}[2]{#2}
\providecommand{\BIBentrySTDinterwordspacing}{\spaceskip=0pt\relax}
\providecommand{\BIBentryALTinterwordstretchfactor}{4}
\providecommand{\BIBentryALTinterwordspacing}{\spaceskip=\fontdimen2\font plus
\BIBentryALTinterwordstretchfactor\fontdimen3\font minus
  \fontdimen4\font\relax}
\providecommand{\BIBforeignlanguage}[2]{{%
\expandafter\ifx\csname l@#1\endcsname\relax
\typeout{** WARNING: IEEEtran.bst: No hyphenation pattern has been}%
\typeout{** loaded for the language `#1'. Using the pattern for}%
\typeout{** the default language instead.}%
\else
\language=\csname l@#1\endcsname
\fi
#2}}
\providecommand{\BIBdecl}{\relax}
\BIBdecl

\bibitem{NetCod2011}
A.~{Kermarrec}, N.~{Le Scouarnec}, and G.~{Straub}, ``{Repairing Multiple
  Failures with Coordinated and Adaptive Regenerating Codes},'' in
  \emph{Network Coding (NetCod), 2011 International Symposium on}, July 2011,
  pp. 1--6, \url{http://dx.doi.org/10.1109/ISNETCOD.2011.5978920}.

\bibitem{Dabek2001}
F.~Dabek, F.~Kaashoek, D.~Karger, R.~Morris, and I.~Stoica, ``{Wide-area
  Cooperative Storage with CFS},'' in \emph{SOSP}, 2001.

\bibitem{Rhea2003}
S.~Rhea, P.~Eaton, D.~Geels, H.~Weatherspoon, B.~Zhao, and J.~Kubiatowicz,
  ``{Pond: the OceanStore Prototype},'' in \emph{FAST}, 2003.

\bibitem{Ghemawat2003}
S.~Ghemawat, H.~Gobioff, and S.-T. Leung, ``{The Google File System},'' in
  \emph{SOSP}, 2003.

\bibitem{Bhagwan2004}
R.~Bhagwan, K.~Tati, Y.-C. Cheng, S.~Savage, and G.~M. Voelker, ``{Total
  Recall: System Support for Automated Availability Management},'' in
  \emph{NSDI}, 2004.

\bibitem{Weatherspoon2002}
H.~Weatherspoon and J.~Kubiatowicz, ``{Erasure Coding Vs. Replication: A
  Quantitative Comparison},'' in \emph{IPTPS}, 2002.

\bibitem{Lin2004}
W.~K. Lin, D.~M. Chiu, and Y.~B. Lee, ``{Erasure Code Replication Revisited},''
  in \emph{P2P}, 2004.

\bibitem{Dimakis2007}
A.~G. Dimakis, P.~B. Godfrey, Y.~Wu, M.~O. Wainwright, and K.~Ramchandran,
  ``{Network Coding for Distributed Storage Systems},'' in \emph{INFOCOM},
  2007.

\bibitem{Dimakis2010}
------, ``{Network Coding for Distributed Storage Systems},'' \emph{IEEE
  Transactions On Information Theory}, vol.~56, pp. 4539--4551, 2010.

\bibitem{Datta2006}
A.~Datta and K.~Aberer, ``{Internet-scale storage systems under churn -- A
  Study of steady-state using Markov models},'' in \emph{P2P}, 2006.

\bibitem{Dalle2009}
O.~Dalle, F.~Giroire, J.~Monteiro, and S.~Pérennes, ``{Analysis of Failure
  Correlation Impact on Peer-to-Peer Storage Systems},'' in \emph{P2P}, 2009.

\bibitem{Hu2010}
Y.~Hu, Y.~Xu, X.~Wang, C.~Zhan, and P.~Li, ``{Cooperative Recovery of
  Distributed Storage Systems from Multiple Losses with Network Coding},''
  \emph{IEEE Journal on Selected Areas in Communications}, vol.~28, pp.
  268--276, 2010.

\bibitem{Wang2010}
X.~Wang, Y.~Xu, Y.~Hu, and K.~Ou, ``{MFR: Multi-Loss Flexible Recovery in
  Distributed Storage Systems},'' in \emph{ICC}, 2010.

\bibitem{Shum2011}
K.~W. Shum, ``{Cooperative Regenerating Codes for Distributed Storage
  Systems},'' in \emph{ICC}, 2011.

\bibitem{Shum2011b}
K.~W. Shum and Y.~Hu, ``{Exact Minimum-Repair-Bandwidth Cooperative
  Regenerating Codes for Distributed Storage Systems},'' in \emph{ISIT}, 2011.

\bibitem{Ahlswede2000}
R.~Ahlswede, N.~Cai, S.-Y. Li, and R.~Yeung, ``{Network Information Flow},''
  \emph{IEEE Transactions On Information Theory}, vol.~46, pp. 1204--1216,
  2000.

\bibitem{Li2003}
S.-Y. Li, R.~Yeung, and N.~Cai, ``{Linear Network Coding},'' \emph{IEEE
  Transactions On Information Theory}, vol.~49, pp. 371--381, 2003.

\bibitem{Koetter2003}
R.~Koetter and M.~Médard, ``{An Algebraic Approach to Network Coding},''
  \emph{IEEE/ACM Transactions on Networking}, vol.~11, pp. 782--795, 2003.

\bibitem{Dimakis2005}
A.~G. Dimakis, V.~Prabhakaran, and K.~Ramchandran, ``{Ubiquitous Access to
  Distributed Data in Large-Scale Sensor Networks Through Decentralized Erasure
  Codes},'' in \emph{IPSN}, 2005.

\bibitem{Dimakis2006}
------, ``{Decentralized Erasure Codes for Distributed Networked Storage},'' in
  \emph{Joint special issue, IEEE/ACM Transactions on Networking and IEEE
  Transactions on Information Theory}, 2006.

\bibitem{Kamra2006}
A.~Kamra, V.~Misra, J.~Feldman, and D.~Rubenstein, ``{Growth Codes: Maximizing
  Sensor Network Data Persistence},'' in \emph{SIGCOMM}, 2006.

\bibitem{Lin2007}
Y.~Lin, B.~Li, and B.~Liang, ``{Differentiated Data Persistence with Priority
  Random Linear Codes},'' in \emph{ICDCS}, 2007.

\bibitem{Ho2006}
T.~Ho, M.~Médard, R.~Koetter, D.~Karger, M.~Effros, J.~Shi, and B.~Leong, ``{A
  Random Linear Network Coding Approach to Multicast},'' \emph{IEEE Transaction
  on Information Theory}, vol.~52, pp. 4413--4430, 2006.

\bibitem{Dimakis2010b}
A.~G. Dimakis, K.~Ramchandran, Y.~Wu, and C.~Suh, ``{A Survey on Network Codes
  for Distributed Storage},'' \emph{The Proceedings of the IEEE}, vol.~99, pp.
  476--489, 2010.

\bibitem{Wu2009}
Y.~Wu and A.~G. Dimakis, ``{Reducing Repair Traffic for Erasure Coding-based
  Storage via Interference Alignement},'' in \emph{ISIT}, 2009.

\bibitem{Duminuco2009}
A.~Duminuco and E.~Biersack, ``{A Pratical Study of Regenerating Codes for
  Peer-to-Peer Backup Systems},'' in \emph{ICDCS}, 2009.

\bibitem{Rashmi2009}
K.~V. Rashmi, N.~B. Shah, P.~V. Kumar, and K.~Ramchandran, ``{Explicit
  Construction of Optimal Exact Regenerating Codes for Distributed Storage},''
  in \emph{Allerton Conference on Control, Computing, and Communication}, 2009.

\bibitem{Shah2010}
N.~B. Shah, K.~Rashmi, P.~V. Kumar, and K.~Ramchandran, ``{Explicit Codes
  Minimizing Repair Bandwidth for Distributed Storage},'' in \emph{ITW}, 2010.

\bibitem{Suh2010}
C.~Suh and K.~Ramchandran, ``{Exact Regeneration Codes for Distributed Storage
  Repair Using Interference Alignment},'' in \emph{ISIT}, 2010.

\bibitem{Tamo2012}
I.~Tamo, Z.~Wang, and J.~Bruck, ``{Access vs. Bandwidth in Codes for
  Storage},'' in \emph{ISIT}, 2012.

\bibitem{Rashmi2011}
K.~V. Rashmi, N.~B. Shah, and P.~V. Kumar, ``{Optimal Exact-Regenerating Codes
  for Distributed Storage at the MSR and MBR Points via a Product-Matrix
  Construction},'' \emph{IEEE Transaction on Information Theory}, vol.~57, pp.
  5227--5239, 2011.

\bibitem{Shah2012}
N.~B. Shah, K.~Rashmi, P.~V. Kumar, and K.~Ramchandran, ``{Distributed Storage
  Codes with Repair-by-Transfer and Non-achievability of Interior Points on the
  Storage-Bandwidth Tradeoff},'' \emph{Transaction on Information Theory},
  vol.~58, pp. 1837--1852, 2012.

\bibitem{Suh2011}
C.~Suh and K.~Ramchandran, ``{Exact-Repair MDS code construction using
  interference alignment},'' \emph{IEEE Transactions On Information Theory},
  vol.~57, pp. 1425--1442, 2011.

\bibitem{Cadambe2011b}
V.~R. Cadambe, C.~Huang, J.~Li, and S.~Mehotra, ``{Polynomial Length MDS Codes
  with Optimal Repair in Distributed Storage Systems},'' in \emph{Proceedings
  of the 45th Asilomar Conference on Signal, Systems and Computing}, 2011.

\bibitem{Tamo2011}
I.~Tamo, Z.~Wang, and J.~Bruck, ``{Zigzag Codes: MDS Array Codes with Optimal
  Rebuilding},'' \emph{IEEE Transaction on Information Theory}, vol.~59, pp.
  1597--1616, 2013.

\bibitem{ElRouayheb2010}
S.~{El Rouayheb} and K.~Ramchandran, ``{Fractional Repetition Codes for Repair
  in Distributed Storage Systems},'' in \emph{Allerton Conference on Control,
  Computing, and Communication}, 2010.

\bibitem{Shah2010b}
N.~B. Shah, K.~Rashmi, P.~V. Kumar, and K.~Ramchandran, ``{Interference
  Alignement in Regenerating Codes for Distributed Storage: Necessity and Code
  Constructions},'' \emph{Transaction on Information Theory}, vol.~58, pp.
  2134--2158, 2012.

\bibitem{Cadambe2010}
V.~R. Cadambe, S.~A. Jafar, and H.~Maleki, ``{Distributed Data Storage with
  Minimum Storage Regenerating Codes - Exact and Functional Repair are
  Asymptotically Equally Efficient},'' in \emph{WiNC}, 2010.

\bibitem{Suh2010a}
C.~Suh and K.~Ramchandran, ``{On the Existence of Optimal Exact-Repair MDS
  Codes for Distributed Storage},'' \emph{ArXiv e-prints}, 2010,
  {arXiv:1004.4663}.

\bibitem{Cadambe2011}
V.~R. Cadambe, S.~A. Jafar, C.~Huang, and J.~Li, ``{Optimal Repair of MDS Codes
  in Distributed Storage via Subspace Interference Alignement},'' in
  \emph{ISIT}, 2011.

\bibitem{Papailiopoulos2011}
D.~S. Papailiopoulos, A.~G. Dimakis, and V.~R. Cadambe, ``{Repair Optimal
  Erasure Codes through Hadamard Designs},'' in \emph{Allerton Conference on
  Control, Computing, and Communication}, 2011.

\bibitem{LeScouarnec2012a}
N.~{Le Scouarnec}, ``{Exact Scalar Minimum Storage Coordinated Regenerating
  Codes},'' in \emph{ISIT}, 2012.

\bibitem{Chen2013}
J.~Chen and K.~W. Shum, ``{Repairing Multiple Failures in the Suh-Ramchandran
  Regenerating Codes},'' \emph{ArXiv e-prints}, pp. 1--5, 2013,
  {arXiv:1302.1256}.

\bibitem{Jiekak2012}
S.~Jiekak and N.~{Le Scouarnec}, ``{CROSS-MBCR: Exact Minimum Bandwidth
  Coordinated Regenerating Codes},'' in \emph{ISIT - Recent Result Poster
  Session}, 2012, {arXiv:1207.0854}.

\bibitem{Li2010}
J.~Li, S.~Yang, X.~Wang, and B.~Li, ``{Tree-structured Data Regeneration in
  Distributed Storage Systems with Regenerating Codes},'' in \emph{INFOCOM},
  2010.

\bibitem{Duminuco2008}
A.~Duminuco and E.~Biersack, ``{Hierarchical Codes: How to Make Erasure Codes
  Attractive for Peer-to-Peer Systems},'' in \emph{P2P}, 2008.

\bibitem{Oggier2010}
F.~Oggier and A.~Datta, ``{Self-repairing Homomorphic Codes for Distributed
  Storage Systems},'' in \emph{INFOCOM}, 2011.

\bibitem{Papailiopoulos2012a}
D.~S. Papailiopoulos and A.~G. Dimakis, ``{Locally Repairable Codes},'' in
  \emph{ISIT}, 2012.

\bibitem{Papailiopoulos2012}
D.~S. Papailiopoulos, J.~Luo, A.~G. Dimakis, C.~Huang, and J.~Li, ``{Simple
  Regenerating Codes: Network Coding for Cloud Storage},'' in \emph{INFOCOM},
  2012.

\bibitem{Oggier2011}
F.~Oggier and A.~Datta, ``{Self-Repairing Codes for Distributed Storage - A
  Projective Geometric Construction},'' in \emph{ITW}, 2011.

\bibitem{Huang2007}
C.~Huang, M.~Chen, and J.~Li, ``{Pyramid Codes: Flexible Schemes to Trade Space
  for Access Efficiency in Reliable Data Storage Systems},'' in \emph{NCA},
  2007.

\bibitem{Huang2012}
C.~Huang, H.~Simitci, Y.~Xu, A.~Ogus, B.~Calder, P.~Gopalan, J.~Li, and
  S.~Yekhanin, ``{Erasure Coding in Windows Azure Storage},'' in \emph{USENIX
  ATC}, 2012.

\bibitem{Huang2008}
C.~Huang and L.~Xu, ``{STAR: An Efficient Coding Scheme for Correcting Triple
  Storage Node Failures},'' \emph{IEEE Transactions on Computers}, vol.~57, pp.
  889--901, 2008.

\bibitem{Prakash2012}
N.~Prakash, G.~M. Kamath, V.~Lalitha, and P.~V. Kumar, ``{Optimal Linear Codes
  with a Local-Error-Correction Property},'' in \emph{ISIT}, 2012.

\end{thebibliography}
\vfill
\pagebreak
\begin{IEEEbiography}[{\includegraphics[width=1in,height=1.25in,clip,keepaspectratio]{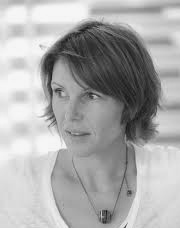}}]{Anne-Marie Kermarrec} is an INRIA researcher.
Before joining INRIA in February 2004, Anne-Marie Kermarrec was with Microsoft Research in Cambridge as a Researcher since March 2000. Before that, she obtained my Ph.D. from the University of Rennes (FRANCE) in October 1996 (thesis). She also spent one year (1996-1997) in the Computer Systems group of Vrije Universiteit in Amsterdam (The Netherlands)  in collaboration with Maarten van Steen and Andrew. S. Tanenbaum and was Assistant Professor at the University of Rennes 1 from 1998 to 2000. She defended her "habilitation à diriger les recherches"  in December 2002 on large-scale application-level multicast.
\end{IEEEbiography}

\begin{IEEEbiography}[{\includegraphics[width=1in,height=1.25in,clip,keepaspectratio]{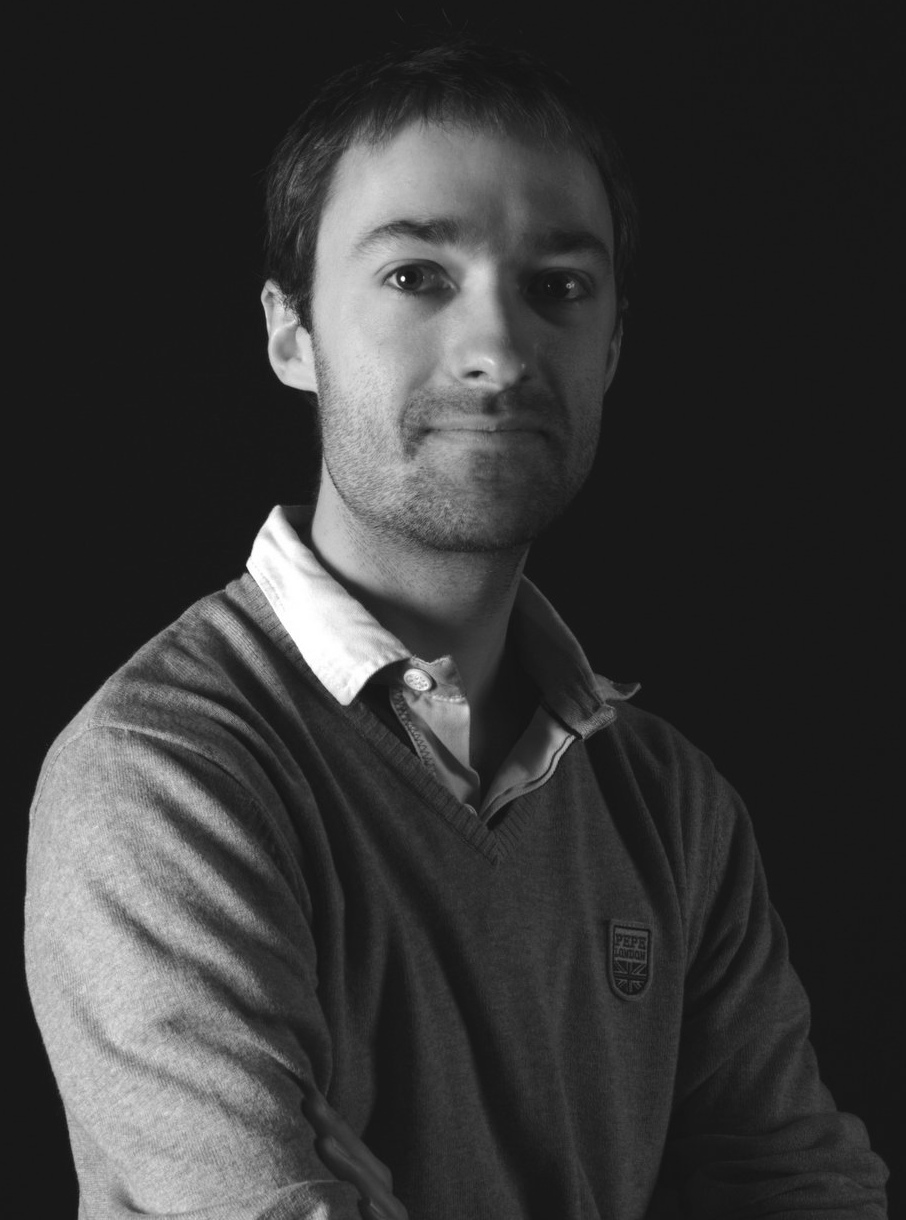}}]{Nicolas Le Scouarnec}
 joined Technicolor in 2007. From 2007 to 2010, He prepared a PhD thesis (Coding for resource optimization in large-scale distributed systems) with Anne-Marie Kermarrec (INRIA Rennes-Bretagne Atlantique) and Mary-Luc Champel (Technicolor). Since 2010, he is now a researcher in Technicolor (Rennes, France) working on distributed storage and cloud computing systems.
\end{IEEEbiography}

\begin{IEEEbiography}[{\includegraphics[width=1in,height=1.25in,clip,keepaspectratio]{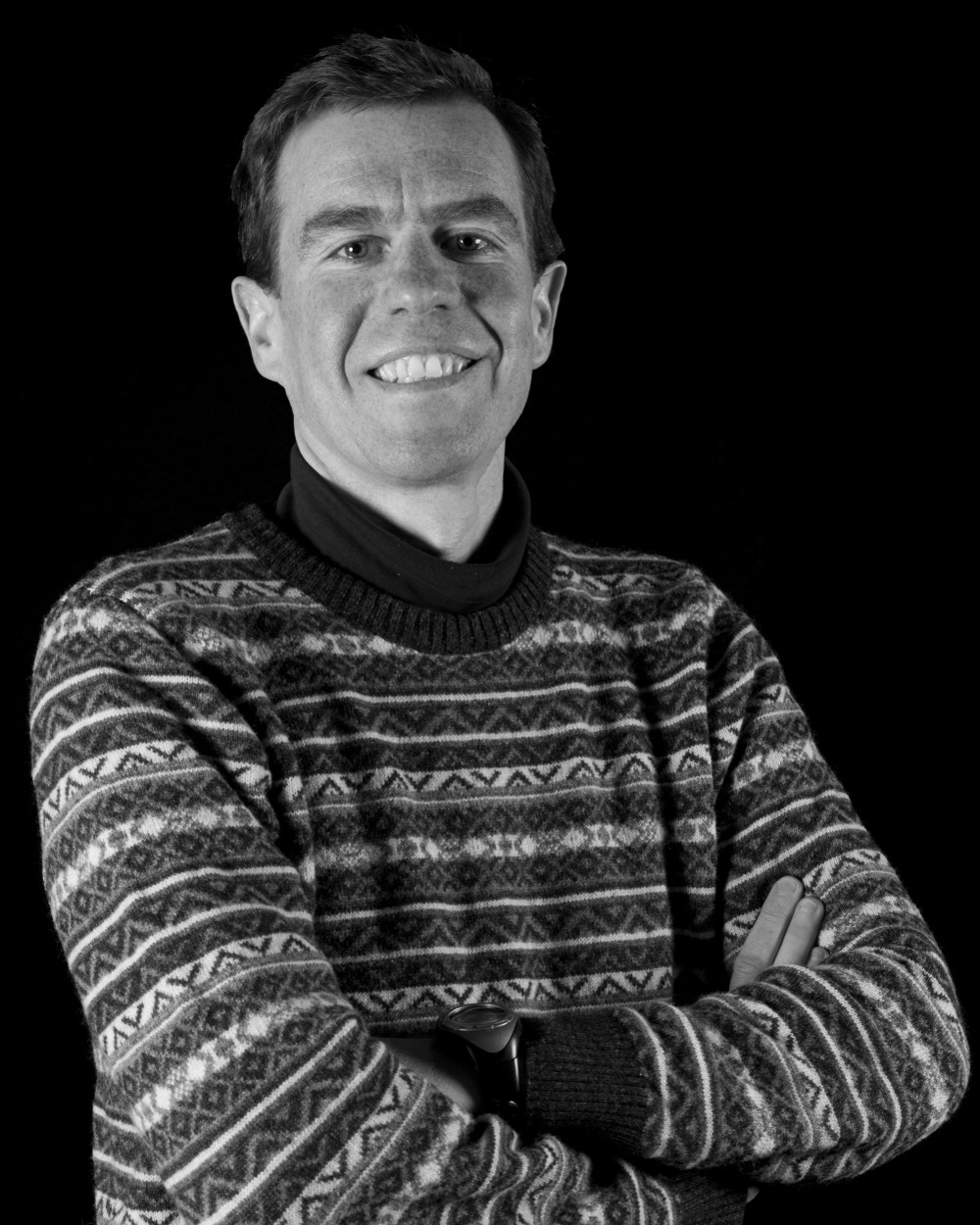}}]
{Gilles Straub} 
was graduated as an Engineer from Ecole Nationale Superieure des Telecom Bretagne in 1991. He started with THOMSON CSF and worked in ATM switching and network adaptations for professional Video equipments. He joined Thomson/Technicolor Research Organization in 1996 and in now Senior Scientist in that company. He actively contributed home networking, wireless and broadband standards, he got the Broadband Forum Circle of Excellence Award in March 2008 for his involvement in TR-135 which is a TR-069 data model of a STB. Since 2008 he is in charge of a work package dealing with distributed media storage. He is co-author of more than 45 patent applications
\end{IEEEbiography}

\vfill

\end{document}